\documentclass[twocolumn,pre,tightenlines,superscriptaddress,floatfix]{revtex4-2}
\usepackage{hyperref}
\usepackage{amsmath,amssymb}
\usepackage[utf8]{inputenc} 			
\usepackage[T1]{fontenc}				
\usepackage{graphicx}
\usepackage{float}
\usepackage[usenames,dvipsnames]{xcolor}
\usepackage{tikz}
\usepackage{ulem}

\newcommand{\eps}{\varepsilon}

\newcommand{\red}[1]{{\color{red} #1}}

\newcommand*\colvec[3][]{
    \begin{pmatrix}\ifx\relax#1\relax\else#1\\\fi#2\\#3\end{pmatrix}
}
\newcommand\bfn{{\bf n}}
\newcommand\C{\mathcal{C}}
\definecolor{vertjul}{rgb}{0,200,23}
\definecolor{violetpale}{HTML}{CCAAFF}
\definecolor{bleupale}{HTML}{CCCCFF}
\definecolor{violetdetonant}{HTML}{FF00FF}
\definecolor{jaunequipete}{HTML}{FFFF66}
\definecolor{darkgreen}{cmyk}{1,0,1,0.1}
\definecolor{trolleygrey}{rgb}{0.5, 0.5, 0.5}
\definecolor{davysgrey}{rgb}{0.33, 0.33, 0.33}
\definecolor{harvardcrimson}{rgb}{0.79, 0.0, 0.09}
\definecolor{lapislazuli}{rgb}{0.15, 0.38, 0.61}
\definecolor{oldlavender}{rgb}{0.47, 0.41, 0.47}
\definecolor{oldrose}{rgb}{0.75, 0.5, 0.51}
\definecolor{ochre}{rgb}{0.8, 0.47, 0.13}
\definecolor{coolblack}{rgb}{0.0, 0.18, 0.39}
\definecolor{persianblue}{rgb}{0.11, 0.22, 0.73}
\definecolor{rossocorsa}{rgb}{0.83, 0.0, 0.0}
\definecolor{carmine}{rgb}{0.59, 0.0, 0.09}
\definecolor{persianred}{rgb}{0.8, 0.2, 0.2}
\definecolor{richelectricblue}{rgb}{0.03, 0.57, 0.82}
\definecolor{cyan(process)}{rgb}{0.0, 0.72, 0.92}
\definecolor{darkgreen}{cmyk}{1,0,1,0.1}

\newcommand\na{{M}}

\begin{document}
\title{Flocking in One Dimension: Asters and Reversals}
\author{Brieuc Benvegnen}
\affiliation{Sorbonne Universit\'e, CNRS, Laboratoire de Physique Th\'eorique de la Mati\`ere Condens\'ee, 75005 Paris, France}

\author{Hugues Chat\'{e}}
\affiliation{Service de Physique de l'Etat Condens\'e, CEA, CNRS Universit\'e Paris-Saclay, CEA-Saclay, 91191 Gif-sur-Yvette, France}
\affiliation{Computational Science Research Center, Beijing 100094, China}
\affiliation{Sorbonne Universit\'e, CNRS, Laboratoire de Physique Th\'eorique de la Mati\`ere Condens\'ee, 75005 Paris, France}

\author{Pavel Krapivsky}
\affiliation{Department of Physics, Boston University, Boston, MA 02215, USA}

\author{Julien Tailleur}
\affiliation{Université Paris Cité, Laboratoire Matière et Systèmes Complexes (MSC), UMR 7057 CNRS,F-75205 Paris, France}

\author{Alexandre Solon}
\affiliation{Sorbonne Universit\'e, CNRS, Laboratoire de Physique Th\'eorique de la Mati\`ere Condens\'ee, 75005 Paris, France}

\date{\today}
\begin{abstract}
  We study the one-dimensional active Ising model in which aligning
  particles undergo diffusion biased by the signs of their spins. The
  phase diagram obtained varying the density of particles, their
  hopping rate and the temperature controlling the alignment shows a
  homogeneous disordered phase but no homogeneous ordered one, as well
  as two phases with localized dense structures.  In the flocking
  phase, large ordered aggregates move ballistically and
  stochastically reverse their direction of motion. In what we termed
  the ``aster'' phase, dense immobile aggregates of opposite magnetization face
  each other, exchanging particles,
   without any net motion of the aggregates. 
   Using a combination of numerical simulations and
  mean-field theory, we study the evolution of the shapes of the flocks, 
  the statistics of their reversal times, and their
  coarsening dynamics.  Solving exactly for the zero-temperature
  dynamics of an aster allows us to understand their coarsening, which
  shows extremal dynamics, while mean-field equations account for
  their shape.
\end{abstract}
\maketitle
\tableofcontents

\section{Introduction}
Active matter often consists of large assemblies of self-propelled
particles.  The variety of collective behaviors that they exhibit has
come under intense
scrutiny~\cite{marchetti_hydrodynamics_2013,bechinger2016active,chate2020dry,o2022time}. This
interest stems from both the pervasiveness of active matter in nature,
from molecular
motors~\cite{schaller2011frozen,sumino_large-scale_2012} to bacterial
swarms~\cite{sokolov_physical_2012,wensink_meso-scale_2012,lopez_turning_2015,liu_self-driven_2019,curatolo2020cooperative}
and animal
groups~\cite{cavagna_scale-free_2010,gautrais_deciphering_2012,attanasi2014collective,poel_subcritical_2022},
and from the possibility to synthesize various kinds of self-propelled
particles, such as chemically propelled Janus
particles~\cite{howse2007self,aranson_active_2013,zhang_janus_2017,ginot2018sedimentation},
rolling
colloids~\cite{bricard2013emergence,geyer_freezing_2019,karani_tuning_2019,liu_activity_2021}
or vibrated macroscopic
objects~\cite{deseigne_collective_2010,kumar_flocking_2014,soni_phases_2020}.

A challenge for physicists is to understand the fundamental
differences between active matter, on the one hand, and passive
systems or other classes of nonequilibrium systems, on the other
hand. The role of fluctuations offers a striking example of such
differences. Indeed, following the introduction of the Vicsek
model~\cite{vicsek_novel_1995}, it was realized by Toner and
Tu~\cite{toner_long-range_1995} that self-propelled particles with a
local ferromagnetic alignment of their direction of motion, a ``flying
XY model'', exhibit an ordered phase of collective motion even in
two-dimensions (2D), whereas such a phase would be destroyed by
fluctuations in a passive system, as requested by the Mermin-Wagner
theorem~\cite{mermin_absence_1966}. Although they do not destroy the
ordered state, density fluctuations are measured to be anomalously
large~\cite{ramaswamy_active_2003,chate_collective_2008,dey_spatial_2012,mahault_quantitative_2019}.
The nature of the transition to the ordered state of collective motion
from a disordered system shows another role of fluctuations. At
mean-field level the transition is predicted to be continuous for many
systems~\cite{ginelli2010relevance,chou2012kinetic,peshkov2012continuous,solon_revisiting_2013}
but becomes discontinuous once fluctuations are accounted
for~\cite{martin_fluctuation-induced_2020}. This translates into large
finite-size effects in simulations: The transition seems continuous in
small systems~\cite{vicsek_novel_1995} but appears
discontinuous~\cite{gregoire_onset_2004,chate_collective_2008}, akin
to a liquid-gas transition, for large-enough system
sizes~\cite{solon_revisiting_2013,solon_phase_2015,martin_fluctuation-induced_2020}.

The effect of fluctuations is most prominent in 1D. For a passive
system with short-range interactions, the Van Hove theorem forbids a
phase transition at non-zero temperature in absence of long-ranged
interactions~\cite{van_hove_sur_1950}. Nonequilibrium systems can
evade this requirement: For example driven diffusive systems show
boundary-induced phase
transitions~\cite{domb_statistical_1995,evans1998phase,kafri2002criterion}.
One-dimensional active systems also order and they do so in a peculiar
way. It was shown in several models of aligning self-propelled
particles that when the noise on alignment is reduced, the system
transitions from a disordered state to a state where a finite fraction
of all particles is contained in a single ordered
aggregate~\cite{czirok_collective_1999,oloan_alternating_1999,raymond_flocking_2006,dossetti_cohesive_2011,solon_revisiting_2013,laighleis_minimal_2018,sakaguchi_flip_2019}. This
``flock'' progressively spreads while  propagating ballistically
until it spontaneously reverses its direction of motion under the
action of fluctuations. Because the time between two reversals
increases only logarithmically with system
size~\cite{oloan_alternating_1999}, there is no symmetry breaking
asymptotically~\cite{solon_revisiting_2013}. The situation is that of
a stochastically switching global order, even in infinite systems.

One-dimensional flocking is also relevant experimentally.  Locusts
marching in a quasi-1D ring-shaped arena order if they are
sufficiently dense and show spontaneous reversals of the global
direction of
motion~\cite{buhl_disorder_2006,yates_inherent_2009}. Recently, the
collective motion of cells confined to 1D racetracks has also been
studied in detail~\cite{bertrand_clustering_2020}.

Here, we analyze the 1D version of the active Ising model (AIM), first
introduced in Ref.~\cite{solon_revisiting_2013}. This lattice model
only involves ferromagnetic alignment and self-propulsion, with no
excluded volume or other interaction. We explore the phase diagram in
terms of the temperature controlling the aligning interaction, the
hopping rate, and the average density of particles. Consistently with
previous studies considering different implementations of the same
ingredients~\cite{czirok_collective_1999,oloan_alternating_1999,dossetti_cohesive_2011,laighleis_minimal_2018,sakaguchi_flip_2019},
we observe a transition from a disordered state to a flocking state
featuring a single moving aggregate, which stochastically reverses its
direction of motion.  Although this transition was reported to be
continuous in early papers on related 1D
models~\cite{czirok_collective_1999,oloan_alternating_1999}, it was
found to be discontinuous in Ref.~\cite{solon_revisiting_2013}, a fact
that is somewhat hidden by reversals. Here we clarify the connection
with the liquid-gas transition happening in
2D~\cite{solon_revisiting_2013,solon_flocking_2015} by showing that
the flocks tend to relax to a phase-separated state but are prevented
to do so by reversals. Moreover, we confirm the logarithmic scaling
with system size of the average reversal time found in
Ref.~\cite{oloan_alternating_1999} for a different 1D flocking model
and track down its origin. At low temperature and/or hopping rate,
when alignment dominates over self-propulsion, we encounter a
different type of structure that we term ``asters''. These asters are
static structures composed of two highly localized peaks of opposite
magnetization that exchange particles back and forth. We first account
for the remarkable stability of these structures by solving exactly
the dynamics of a single aster at zero temperature (this is the
asymptotic configuration then).  We then discuss the
finite-temperature coarsening of asters and show the existence of two
possible scenarios leading either to macroscopic condensation or to an
extensive number of asters possessing a characteristic size. Finally,
we derive the steady-state profile of the asters in the mean-field
approximation, which turns out to be very accurate.

This article is organized as follows: In Sec.~\ref{sec:phases} we first
define the AIM and present the three phases observed in our
simulations (disordered, flocking and aster phases) with phase
diagrams for the main parameters of the model. We then investigate the
two non-trivial phases. In Sec.~\ref{sec:flocks},
we successively look at the shapes of asters and how they evolve in time
(Sec.~\ref{sec:flocks-shape}), the statistics of reversals
(Sec.~\ref{sec:flocks-reversal}), and the two distinct regimes observed
at small and large velocity and/or temperature
(Sec.~\ref{sec:flocks-v}). We then investigate the aster phase in
Sec.~\ref{sec:asters}. We first derive an exact solution for the
dynamics of an aster at zero temperature in Sec.~\ref{sec:asterT0},
before looking at the coarsening dynamics at small-but-finite
temperature in Sec.~\ref{sec:asterTsmall} and investigating the aster
shape in Sec.~\ref{sec:astersProfile}. 

\begin{figure}
    \centering
    \includegraphics[width=0.97\columnwidth]{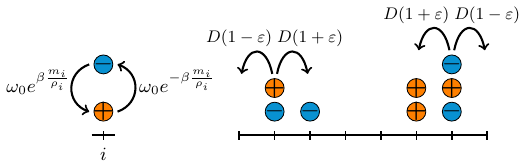}

    \caption{Sketch of the two possible actions, spin-flipping and
      hopping, and their associated rates.}
    \label{fig:rates}
\end{figure}

\section{Model and its phases}
\label{sec:phases}

\subsection{1D active Ising model}

Following Refs.~\cite{solon_revisiting_2013,solon_flocking_2015}, we
consider $N$ particles on a ring of $L$ sites.  There is no
excluded-volume interaction so that each site can accommodate an
arbitrary number of particles. Each particle carries an Ising spin
$s=\pm 1$ and undergoes biased diffusion: A particle with spin $s$
hops to the next site on its right (resp. left) at rate $D(1+s\eps)$
(resp. $D(1-s\eps)$).  The parameter $\eps\in [0,1]$ controls the
asymmetry between the passive limit $\eps=0$ and the fully asymmetric
hopping $\eps=1$, while $D$ controls the overall hopping frequency. In
average, a particle thus moves at a speed $v=2D\eps a$ in the
direction set by the sign of its spin, $a$ being the lattice
spacing. Finally, on a site $i$ occupied by $n_i^+$ and $n_i^-$
particles with spins $+1$ and $-1$, respectively, a particle of spin
$s$ flips at a rate
\begin{equation}\label{eq:flipping-rates}
  W(s\to -s)=\omega_0 \, e^{-\beta s \frac{m_i}{\rho_i}}\;,
\end{equation}
where $\rho_i=n_i^++n_i^-$ and $m_i=n_i^+-n_i^-$ are the local density
and magnetization. This rate is chosen such that the flipping dynamics
of each site $i$ is independent and corresponds to that of a
fully-connected on-site Ising model with Hamiltonian $-m_i^2/(2\rho_i)$
undergoing an equilibrium dynamics at inverse temperature
$\beta=1/T$. The two actions (hopping and flipping spin) and the
associated rates are depicted in Fig.~\ref{fig:rates}. Note that the
model is out of equilibrium even at $\eps=0$ since the symmetric
hopping dynamics is insensitive to the changes of the total
Hamiltonian $H=-\sum_i m_i^2/(2\rho_i)$~\footnote{In this limit, the
  hopping dynamics can be seen as an equilibrium dynamics at infinite
  temperature, while the flipping dynamics occurs at temperature
  $T$.}.

In the rest of the paper, we choose without loss of generality
$\omega_0=a=1$ thus fixing the time and space units. We study the
system as a function of the parameters $\beta$ (or $T$), $D$, $\eps$
and the average density $\rho_0=N/L$. Our simulations relied either on
discrete time steps with random sequential updates or on an exact
continuous-time Monte Carlo algorithm. The efficiencies of both
schemes depend on the temperature and phases under study, as detailed
in Appendix~\ref{sec:numerics}.

\subsection{The three phases}
\begin{figure}
    \centering
    \includegraphics[width=0.48\columnwidth]{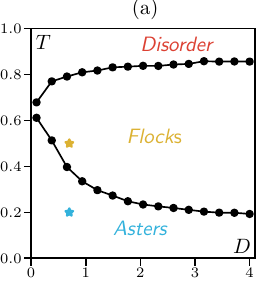}
    \includegraphics[width=0.48\columnwidth]{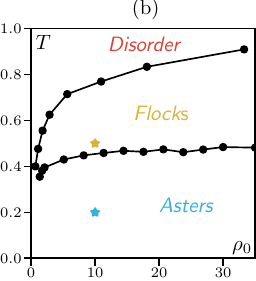}
    \caption{Phase diagrams in the (hopping-rate, temperature) plane
      at $\rho_0=10$ and $\eps=0.7$ (left) and (density, temperature)
      plane at $D=0.5$ and $\eps=0.7$ (right). The stars indicate the
      parameters used in Fig.~\ref{fig:phases} to illustrate the
      phases. The line separating the disordered and flocking phases
      in the ($\rho_0, T$) plane is set by the density $\rho_g(T)$ of
      the gas in the flocking states at temperature $T$ (see
      Sec.~\ref{sec:LG}). The other lines are determined using the
      threshold $\langle |\tilde m|\rangle=0.05$ on the time-averaged
      magnetization per particle, which is non-zero only in the
      flocking phase.  Data obtained in a system of size $L=500$;
      increasing system size displaces the lines only within the size
      of the symbols.}
    \label{fig:phase-diagrams}
\end{figure}

Looking at the phase diagrams in the $D-T$ and $\rho_0-T$ planes shown
in Fig.~\ref{fig:phase-diagrams} (and in the qualitatively similar
$\eps-T$ diagram not shown), we see that, as expected, 
the system is disordered at high temperature. Density and magnetization are homogeneous
and fluctuate, respectively, around $\rho_0$ and $0$. 

\begin{figure}
    \centering
    \includegraphics[width=0.98\columnwidth]{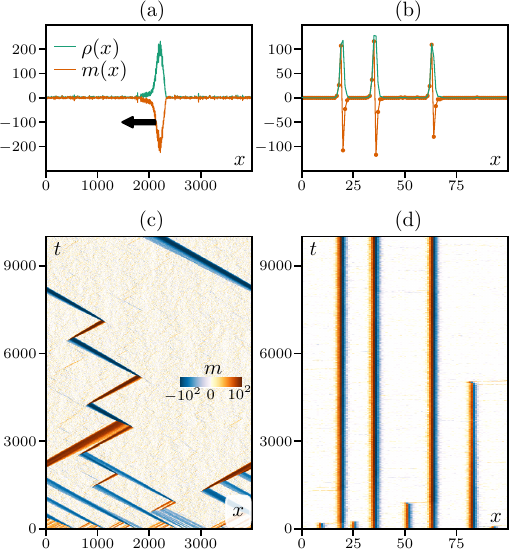}
    \caption{{\bf (a):} The magnetization per spin $\tilde m$, is
      non-vanishing only in the flocking phase. {\bf (b):} The
      fraction of particles $\phi$ contained in asters is
      non-vanishing only in the aster phase. {\bf (c-d):}
      Instantaneous density and magnetization fields at the final time
      $t=10^4$ of the space-time diagrams shown in (e-f). {\bf (e-f):}
      Space-time diagrams in the flocking (e) and aster (f) phases.
      The parameters, indicated by stars in the phase diagrams of
      Fig.~\ref{fig:phase-diagrams}, are $\rho_0=10$, $\eps=0.7$, $D =
      0.5$ and $\beta=2,\,5$ for the F and A phases,
      respectively. The system sizes $L=4000$ (F phase) and $L=100$ (A
      phase) were chosen for legibility. Simulations are started from
      a homogeneous disordered initial condition. The runs in (a,b)
      for the A and F phases are the same as in (e,f).}
    \label{fig:phases}
\end{figure}

Decreasing temperature, the system reaches a flocking phase which
consists, at long times, of a single dense ordered aggregate moving
ballistically in a disordered gas, as illustrated on the snapshot
shown in Fig.~\ref{fig:phases}c. These ``flocks'' have a rather
complex dynamics: They slowly but regularly widen as they travel, and
undergo stochastic reversals during which they ``regroup'' into a very
thin condensate and their magnetization changes sign. These dynamics,
together with the coarsening leading to a single aggregate, are
displayed in the space-time diagram of Fig.~\ref{fig:phases}e, and form
the topic of Sec.~\ref{sec:flocks}. Note that transient, finite-size
aggregates can be observed to move ballistically in the gas region
surrounding the main aggregate (see Fig.~\ref{fig:phases}e).

Finally, at lower temperatures, the system exhibits what we have
called asters, that are illustrated in the snapshot shown in
Fig.~\ref{fig:phases}d: Sharp peaks of positive and negative
magnetizations, spread over a few sites, face each other. These
structures are long-lived, despite the absence of repulsive
interactions. As shown on the space-time plot Fig.~\ref{fig:phases}f,
asters can dissolve on long time scales, leading to coarsening. Our
study of this phase is in Sec.~\ref{sec:asters}.

To distinguish between the three phases, we introduce two order
parameters, the magnetization per particle
$\tilde m=\sum_{i=1}^N s_i /N$, which is non-zero at long times only
in the flocking phase where it alternates between a positive and a
negative value because of reversals, and the fraction $\phi$ of
particles contained in asters. Our criterion to detect an aster is to
find two adjacent sites with magnetization of opposite sign and
density larger than $2\rho_0$. (This value is chosen such that asters
are robustly detected while not erroneously counting fluctuations as
asters. Our results do not depend on the precise criterion used.)
The two order parameters are shown as a function of time in the
flocking and aster phases (Fig.~\ref{fig:phases}a-b).

\subsection{Fate of the liquid-gas transition scenario}
\label{sec:LG}

In 2D, the transition to collective motion in polar flocking models
has been shown to be akin to a liquid-gas transition between a
disordered gas and a polar ordered
liquid~\cite{solon_revisiting_2013,solon_flocking_2015}. The form of
the phase coexistence depends on the symmetry of the spins: for the
Vicsek model with continuous spins one observes microphase
separation with an extensive number of dense ordered traveling bands having a
characteristic size, while in the AIM, which has a discrete spin
symmetry, one obtains phase separation between two macroscopic
domains~\cite{solon_phase_2015}.

In the same way, in 1D one observes the coexistence of an ordered
flock with a disordered gas. Although, because of fluctuations it can
easily be mistaken for a continuous
transition~\cite{czirok_collective_1999,oloan_alternating_1999}, the
transition between the disordered phase and the flocking state also
shows a phase-separation scenario. In particular, the transition
exhibits metastability and the telltale negative peaks in the Binder
fourth-order cumulant for large (but finite) systems which arise from
discontinuous nucleation
events~\cite{binder_finite-size_1984,solon_revisiting_2013}. In
addition, the density in the gas is independent of the average density
$\rho_0$ in the system.

\begin{figure}
    \centering
    \includegraphics[width=0.49\columnwidth]{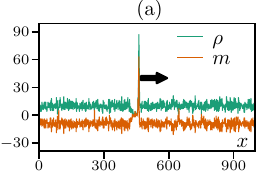}
    \includegraphics[width=0.49\columnwidth]{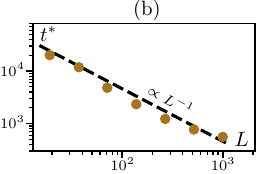}
    \caption{{\bf Left:} Instantaneous profile showing a fluctuation
      propagating through the homogeneous ordered phase. {\bf Right:}
      Average life time $t^*$ of the metastable homogeneous ordered
      phase. Parameters: $L=1000$ (left), $\beta=2$, $\eps=1$, $D =
      0.5$ and $\rho_0=10$.}
    \label{fig:metastab}
\end{figure}

However, looking at the phase diagram in the ($\rho_0,T$) plane
(Fig.~\ref{fig:phase-diagrams}, right), a second transition line to a
homogeneous ordered liquid is conspicuously absent. This is because,
contrary to what happens in 2D, fluctuations destroy the homogeneous
ordered phase in 1D. Indeed, as in Ref.~\cite{raymond_flocking_2006},
we see that, if we prepare the system in this state, it is metastable
but eventually gets destabilized by a fluctuation. As shown in the
left panel of Fig.~\ref{fig:metastab} (see also Supplementary
  Movie 1), a local fluctuation of the opposite magnetization
propagates through the entire system, until only a localized flock
remains.  Thus, even if the probability of such a fluctuation may be
rare, it increases linearly with system size since it can happen
anywhere in the system. Therefore, we expect that the lifetime of the
ordered phase decreases as $L^{-1}$, which is what we measure (see
Fig.~\ref{fig:metastab}, right). At large enough system size, the
ordered phase is thus destabilized very quickly.

The form of the phase coexistence is also altered by fluctuations in
1D. 
Only the gas density $\rho_g$ is easily defined and
separates the disordered and flocking state in the phase diagram
Fig.~\ref{fig:phase-diagrams} (right). 
Below, we study how flocks tend to relax to
a phase-separated state as in 2D but are
prevented to do so by reversals.

\section{The flocks and their reversals}
\label{sec:flocks}

\begin{figure}
    \centering
    \includegraphics[width=0.9\columnwidth]{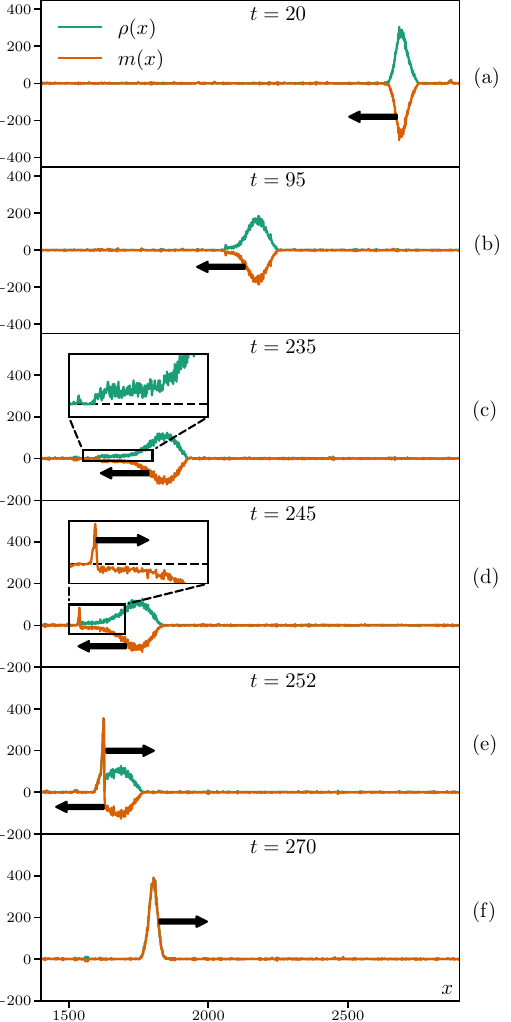}
    \caption{Instantaneous density (green) and magnetization (orange)
      profiles during the spreading and the reversal of a flock. An
      initially peaked aggregate (a) spreads while propagating
      (b). After some time, it develops a protrusion at the leading
      edge (c). A spontaneous fluctuation at the front (d) can
      propagate inside the flock and progressively flip all the
      particle orientations (e). Finally, just after the reversal, one
      is left with a flock with the same shape as in (a) (which was
      taken just after the previous reversal), albeit with a reversed
      magnetization. Time is counted from the previous
      reversal. Parameters: $L=4000$, $\beta=2$, $\eps=1$, $D = 0.5$,
      $\rho_0=10$.}
    \label{fig:snap-reversal}
\end{figure}

In the steady state of the flocking phase, the system typically presents a single ordered aggregate: the flock. 
As already visible in Fig.~\ref{fig:phases}e
and detailed in Fig.~\ref{fig:snap-reversal} (see also
  Supplementary Movie 2), this flock moves ballistically but its
shape evolves continuously: initially narrow and sharp, it
progressively spreads. After some time, a protrusion with a
well-defined density corresponding to the (metastable) homogeneous
liquid phase begins to develop at the leading edge, growing out of the
main peak.  Once the flock is sufficiently spread, a fluctuation
that flips the first few sites at the front can become large enough
that it flips systematically all the remaining flock particles.  This
leads to a full reversal of the flock, whose dynamics then resets,
starting with the same initial sharply peaked shape but with the
opposite magnetization (compare Fig.~\ref{fig:snap-reversal}a and
(f)).

In this section, we study quantitatively this dynamics. We first look
at the evolution of the shape of the flock (Sec.~\ref{sec:flocks-shape}), which provides insight into how the
time between two reversals varies with system size
(Sec.~\ref{sec:flocks-reversal}) and with other parameters
(Sec.~\ref{sec:flocks-v}). Finally we  study the coarsening process following
a quench from the disordered state (Sec.~\ref{sec:flocks-coarsening}).

\subsection{The shape of a flock}
\label{sec:flocks-shape}

\begin{figure*}
    \centering
    \includegraphics[width=0.32\textwidth]{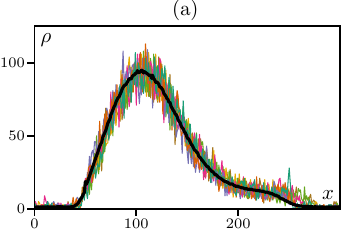}
    \includegraphics[width=0.32\textwidth]{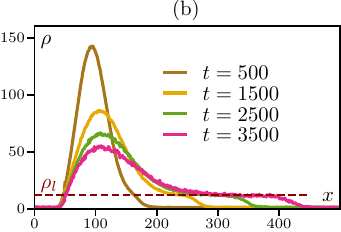}
    \includegraphics[width=0.32\textwidth]{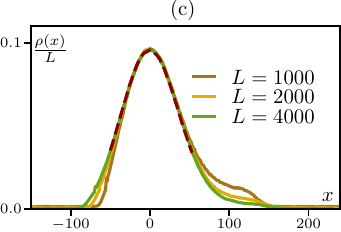}\\
    \includegraphics[width=0.32\textwidth]{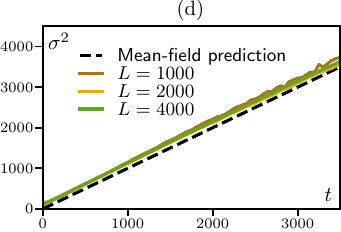}
    \includegraphics[width=0.32\textwidth]{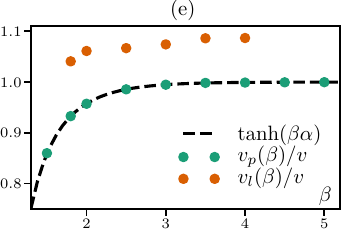}
    \includegraphics[width=0.32\textwidth]{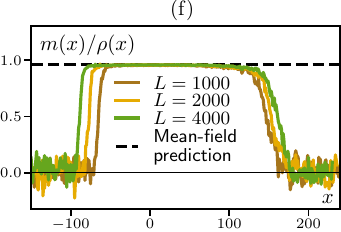}
    \caption{{\bf (a):} Seven representative instantaneous density
      profiles (colors) superimposed to the average profile (thick
      black line). {\bf (b):} Evolution of the averaged density
      profile as a function of time. The red dashed line is a fit to
      the density $\rho_\ell$ of the protrusion. {\bf (c):} Average
      density profiles at $t=1200$ showing that the main peak is
      extensive in system size. The red dashed line is a Gaussian fit
      used to compute the variance $\sigma^2$ in (d). {\bf (d):}
      Variance of the main peak. The dashed line is the mean-field
      prediction $\sigma^2= 2D t$. {\bf (e):} Speeds $v_p$ of the main
      peak (computed as the speed of the maximum) and $v_\ell$ of the
      protrusion (computed as the speed of the point at
      $\rho=\rho_\ell/2$ in the average profiles) divided by the
      self-propulsion speed $v=2 D \varepsilon$. The dashed line is
      the mean-field prediction.  {\bf (f):} Average polarization
      profile $m(x)/\rho(x)$. The dashed line is the mean-field prediction,
      solution of $m/\rho=\tanh(\beta m/\rho)$.  Parameters: $\rho_0 =
      10$, $\beta = 2$, $\varepsilon=1$, $D=0.5$, $t=1000$ (a,c,e,f),
      $L = 1000$ (a,b), $L=5000$ (e).}
    \label{fig:shape}
\end{figure*}

To separate the trend from fluctuations in the evolution of the flock
shapes, we average them over many realizations and construct the
average flock shape at time $t$ after their last reversal. To do so,
we define an origin of time for each reversal and spatially align
flocks of similar age $t$. As shown in Fig.~\ref{fig:phases}a, the
magnetization per spin $\tilde{m}$ flips from a well defined value
$\tilde m_0$ to $-\tilde m_0$ during a reversal. We define the time
origin of each reversal as the time when
$\tilde m(t)=-0.7 \tilde m_0$. (The 0.7 factor works well, but other
values are of course possible.)  Flocks of the same age are localized
at different positions on the lattice.  A simple idea to align them so
as to be able to average their shape would be to align their densest
sites to try and match their peaks. However, this is very noisy and
leads to large fluctuations that artificially smear out the flock
shapes. Instead, we use the rear edges of the flocks, which are always
very sharp. (The detailed aligning procedure is described in
Appendix~\ref{sec:numerics-align}.)  This leads to satisfying results:
As shown in Fig.~\ref{fig:shape}a, flocks superimpose with little
spread along most of their profiles, allowing to extract meaningful
average shapes. Note however that the precise position of the leading
edge fluctuates, leading to an average profile whose leading edge is
smoother than that of the instantaneous ones.  This confirms clearly
the two features mentioned earlier: the flock contains a main peak
which spreads continuously and, after some time, a protrusion with
constant density and magnetization which develops at its leading edge
(see Fig.~\ref{fig:shape}b). We now discuss both of these features.

As shown in Fig.\ref{fig:shape}c, the main peak of the flock at a
given age is proportional to the system size and contains a
macroscopic fraction of the particles. The top of the peak is well
approximated by a Gaussian with a variance $\sigma^2$ that is
independent of system size and grows linearly in time
(Fig.~\ref{fig:shape}d). In addition, the peak propagates at a
constant speed $v_p$, defined as the speed of the maximum of the
density profile, which is smaller than the speed $v$ of individual
particles (Fig.~\ref{fig:shape}e).

Since the peak contains a high density of particles, we expect it to
be well described by a mean-field theory that we now construct
starting from the microscopic dynamics, following
Ref.~\cite{solon_flocking_2015}.  To do so, we first write exact
equations for the average density and magnetization
$\langle \rho_i\rangle$ and $\langle m_i\rangle$ on site $i$, the
average being over realizations of the stochastic microscopic
dynamics. This gives
\begin{align}
  \partial_t \langle \rho_i\rangle&\!=\!D\langle \rho_{i+1}+\rho_{i-1}-2\rho_i\rangle-\tfrac{v}{2}\langle m_{i+1}-m_{i-1}\rangle   \label{eq:hydro-rho}\\
  \partial_t \langle m_i\rangle&\!=\!D\langle m_{i+1}+m_{i-1}-2m_i\rangle-\tfrac{v}{2}\langle \rho_{i+1}-\rho_{i-1}\rangle \nonumber\\ &+2\Big\langle \rho_i\sinh\left[\beta \tfrac{m_i}{\rho_i}\right] -m_i\cosh\left[\beta \tfrac{m_i}{\rho_i}\right]\Big\rangle \label{eq:hydro-m}
\end{align}
These equations are not closed since Eq.~(\ref{eq:hydro-m}) involves
the average of a nonlinear term. However, in the mean-field
approximation where fluctuations and correlations are neglected,
$\langle f(x)\rangle=f(\langle x\rangle)$ for any function $f$. Taking
in addition the continuous limit,
Eqs.~(\ref{eq:hydro-rho})-(\ref{eq:hydro-m}) rewrite
\begin{align}
  \label{eq:MF2-rho}
  \partial_t \rho& \!=\! D\partial_x^2 \rho -v\partial_x m \\
  \label{eq:MF2-m}
  \partial_t m& \!=\! D\partial_x^2 m \!-\! v\partial_x \rho\!-\!2m \cosh\left[\beta\tfrac{m}{\rho}\right]\!+\!2\rho\sinh\left[\beta\tfrac{m}{\rho}\right]\!. 
\end{align}
with $\rho(x=ia)=\langle \rho_i\rangle$ and similarly for $m$.

Contrary to the density field, $m(x,t)$ is a fast mode which relaxes
rapidly to a value that makes the right-hand-side of
Eq.~\eqref{eq:MF2-m} vanish. To leading order in a gradient expansion,
this amounts to requiring that the interaction term in
Eq.~(\ref{eq:MF2-m}) vanishes, i.e. that $m/\rho=\tanh(\beta
m/\rho)$. This thus predicts $\alpha=m/\rho$ to be the solution of
$\alpha=\tanh(\beta\alpha)$, which is verified numerically in
Fig.~\ref{fig:shape}f. Using $m(x)=\alpha \rho(x)$ in
Eq.~(\ref{eq:MF2-rho}), we obtain a diffusion-drift approximation for
the density field:
\begin{equation}
  \label{eq:MF-Pdd}
  \partial_t \rho \!=\!D \partial_x^2 \rho-v\tanh(\beta\alpha)\partial_x \rho.
\end{equation}
The diffusion and drift coefficients read from Eq.~(\ref{eq:MF-Pdd})
both match numerical measurements as seen in Fig.~\ref{fig:shape}d-e.

\begin{figure}
    \centering
    \includegraphics[width=0.48\columnwidth]{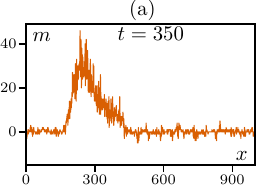}
    \includegraphics[width=0.48\columnwidth]{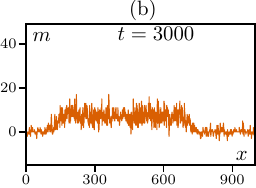}
    \caption{Instantaneous density profile at times $t=350$ (left) and
      $t=3000$ (right) after the last reversal. The profile relaxes to
      a long-lived phase-separated state, whose shape is stationary
      (until the next reversal occurs). Parameters: $\rho_0=5$,
      $L=1000$, $\beta=2$, $\eps=1$ and $D=2$.}
    \label{fig:phase-separation}
\end{figure}

Let us now turn to the protrusion appearing at the front of the
flock. Since reversals start at the leading edge, this protrusion is
expected to play an important role in the dynamics. Contrary to the
main peak, it has a fixed height (in density and magnetization)
independent of time and system size (not shown). Its velocity results
both from the persistent hop of the particles and from the recruitment
of new sites at the leading edge due to the aligning dynamics. This
explains why the leading edge of the protrusion grows with a speed
larger than the individual particle speed $v$
(Fig.~\ref{fig:shape}e), much like for the polar bands of the
coexistence region in 2D~\cite{solon_flocking_2015}. The front speed
is thus also larger than the drift velocity of the peak, which
explains the increase of the protrusion length as time passes. All
this suggests that the flock is trying to relax to the fixed density
$\rho_\ell$ of the homogeneous ordered liquid but is prevented to do
so by the reversals. This is confirmed by choosing parameters such
that the full phase separation can be observed before a reversal
happens. This is achieved in Fig.~\ref{fig:phase-separation} by
increasing $D$; one then observes a relatively long-lived
phase-separated profile.

\subsection{Statistics of reversal times}
\label{sec:flocks-reversal}

\begin{figure}
  \centering
  \begin{minipage}{0.48\columnwidth}
    \includegraphics[width=1\columnwidth]{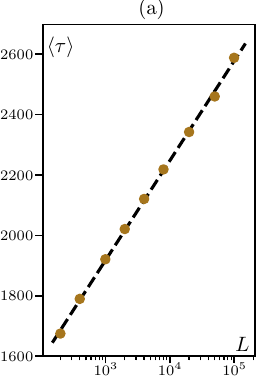}
  \end{minipage}
    \begin{minipage}{0.48\columnwidth}
    \includegraphics[width=1\columnwidth]{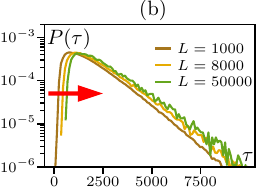}\\
    \includegraphics[width=1\columnwidth]{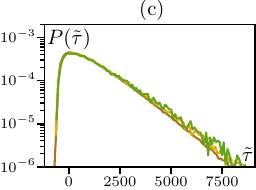}
   \end{minipage}
    \caption{{\bf (a):} Mean time between reversals with a fit to a
      logarithmic function $f(L)=a\log L+b$, leading to $a=139$ and $b=964$
      (dashed line). {\bf (b):} Probability density function of the
      reversal time $\tau$. The red arrow emphasizes the shift of the
      distributions at small $\tau$ responsible for the logarithmic
      scaling of the mean. {\bf (c):} Same as in (b) with an horizontal
      shift $\tilde\tau=\tau-\alpha \log L$ with $\alpha=110$. Parameters:
      $\rho_0 = 10$, $\beta = 2$, $\eps=1$, $D=0.5$.}
    \label{fig:tau-L}
\end{figure}

The average time between two reversals, $\langle \tau \rangle$, shows
a logarithmic increase with system size $\langle \tau \rangle \propto
\log(L)$ (Fig.~\ref{fig:tau-L}a). This is in line with previous
results obtained by O'Loan and Evans for a different
model~\cite{oloan_alternating_1999} (see also \cite{raymond_flocking_2006}), 
and it confirms that no true
symmetry-breaking arises: the system spends finite fractions of time
going to the right, to the left, and in the
reversals~\cite{solon_revisiting_2013}.

To understand the physical origin of this logarithmic scaling, we
consider $P(\tau)$, the distribution of inter-reversal times. 

As shown
in Fig.~\ref{fig:tau-L}b, it has a peak and an approximately
exponential tail with a decay rate that is roughly independent of
$L$. Increasing $L$, the distribution shifts slightly to the right. As
shown in Fig.~\ref{fig:tau-L}c, a horizontal shift by $\alpha \log
L$, with $\alpha$ a constant, collapses reasonably well the
distributions obtained for several values of $L$.

\begin{figure}
    \centering
    \includegraphics[width=0.48\columnwidth]{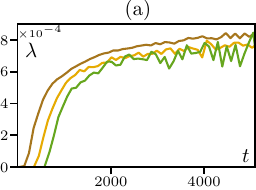}
    \includegraphics[width=0.48\columnwidth]{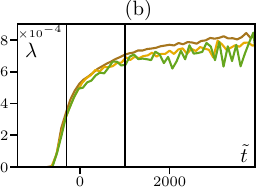}
    \includegraphics[width=0.48\columnwidth]{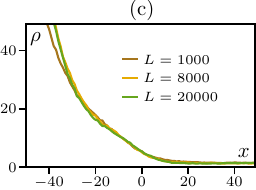}
    \includegraphics[width=0.48\columnwidth]{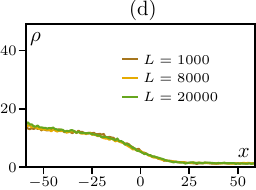}
    \caption{{\bf Top:} Rate of reversal $\lambda(t)$ as a function of
      time since the last reversal {\bf (a)} and with time shifted as
      $\tilde t=t-\alpha\log L$ {\bf (b)}, $\alpha=110$ as in
      Fig.~\ref{fig:tau-L}. {\bf Bottom:} Average profiles at fixed
      $\tilde t$, at $\tilde t=-300$ when the protrusion just begins
      to appear {\bf (c)}, and at $\tilde t=1000$ when it has
      developed ({\bf d}). The corresponding times are indicated by
      vertical lines in panel {(b)}. The profiles are aligned on the
      point where $\rho=5$. Parameters: $\rho_0 = 10$, $\beta = 2$,
      $\eps=1$, $D=0.5$.}
    \label{fig:lambda}
\end{figure}

To better characterize the various processes at play, we compute the
 reversal rate $\lambda(t)$, defined via the probability $\lambda(t)dt$
that a flock of age $t$ reverses within $[t,t+dt]$. This rate
is related to the distribution $P(\tau)$ via the number of flocks $N_f(t)$
that have survived until time $t$ from an initial population of $N_f(0)$:
$\dot N_f(t)=-\lambda(t) N_f(t)$ so that $\lambda=-\dot
N_f/N_f$. In addition $N_f$ itself is related to $P$ through
$N_f(t)/N_f(0)=\int_t^\infty P(\tau)d\tau$, which yields
\begin{equation}
  \label{eq:lambda}
  \lambda(t)=\frac{P(t)}{\int_t^\infty P(\tau)d\tau}\;.
\end{equation}
In physical terms, Eq.~\eqref{eq:lambda} simply states that the
probability that an event happens at time $t$ is the rate of occurence
of this event at time $t$ multiplied by the probability that it did
not happen before: $P(t)=\lambda(t) [1-\int_0^t P(\tau)d\tau]$.
Note that for an exponential distribution, $\lambda$ equals the decay rate of
the exponential.
In Fig.~\ref{fig:lambda}a, we show
$\lambda(t)$ computed using Eq.~\eqref{eq:lambda} for several system
sizes. We see an initial rapid increase in $\lambda$, with a
characteristic time 
that increases with system size, followed by
an approximately flat plateau corresponding to the exponential tail of
$P(\tau)$. As shown in Fig.~\ref{fig:lambda}b, shifting the time by
$\alpha \log L$ provides a good collapse of the curves $\lambda(t-\alpha\log L)$
measured for different system sizes, consistent with what was
reported for $P(\tau-\alpha\log L)$ in Fig.~\ref{fig:tau-L}.

The evolution of $\lambda(t)$ can be related to that of the shape of
the flock. Reversals are initiated at the leading edge of flocks. At
short times, the leading edge is very stiff so that there is little
chance that a spontaneous fluctuation in the gas ahead, whose typical
density is very low, triggers a reversal
(Fig.~\ref{fig:snap-reversal}a). Correspondingly, $\lambda(\tau)\to
0$ as $\tau\to 0$. At longer times two processes take place that make
reversals more likely. First, flocks spread diffusively, due to the
stochastic hopping of the particles. As the leading edge smoothens, it
becomes more and more susceptible to fluctuations.  Then, once the
leading edge has sufficiently spread, its shape becomes compatible with
the development of a liquid phase~\cite{solon_pattern_2015}.  As this
liquid protrusion develops, the shape of its leading front becomes
constant in time (Fig.~\ref{fig:snap-reversal}b).  The probability
that a fluctuation flips the protrusion becomes time-independent,
leading to a plateau value of $\lambda(t)$ at late times, independent
of $L$.

Let us now try to rationalize the scaling form $P(\tau-\alpha\log L)$
and $\lambda(t-\alpha \log L)$ reported in Figs.~\ref{fig:tau-L}
and~\ref{fig:lambda}. We see in Figs.~\ref{fig:lambda}c-d that the
leading edges of the density profiles also collapse under the same
shift.  To account for this, we compute the time $t^*$ it takes for
the profile to reach a given slope $-k^*$ at a given density
$\rho_\ell$:
\begin{equation}
  \label{eq:conditions-k}
  \rho(x,t^*)=\rho_\ell; \quad \partial_x\rho(x,t^*)=-k^*.
\end{equation}
We first focus on the early-time dynamics, where we expect a Gaussian
spreading in the co-moving frame. There, the density profile can be
approximated as
\begin{equation}
  \label{eq:rho-Gaussian}
  \rho(x,t)=\frac{N_0}{\sqrt{2\pi D t}}e^{-\frac{x^2}{4 D t}}\;,
\end{equation}
where $D$ is an effective diffusion coefficient and $N_0=L \rho_0$ is the
number of particles in the aggregate. Denoting by $x^*$ the position
at which the density profile equals $\rho_\ell$, we find that the
slope satisfies $k^*=x^* \rho_\ell/(2 Dt)$ while $x^*$ is given by
\begin{equation}
  x^*=2 \sqrt{D t} \sqrt{\log\frac{N_0}{\rho_\ell}-\frac 1 2 \log (2 \pi D t)}\;.
\end{equation}
\if{At early times and large system sizes,
$\log (2\pi Dt)\ll \log\frac{N_0}{\rho_\ell}\propto \log L$, so that 
$x^* \propto 2\sqrt{Dt \log L}$, and thus $k^*\propto \rho_\ell \sqrt{(\log L)/Dt}$.
In this limit, $k^*$ is a function of $t/\log L$.}\fi
The slope as a function of time then satisfies
\begin{equation}\label{eq:slope}
 k(t)=\rho_\ell \sqrt{\frac{\log \frac{N_0}{\rho_\ell}}{D t} \left(1-\frac 1 2 \frac{\log 2\pi Dt}{\log \frac{N_0}{\rho_\ell}}\right)}.
\end{equation}

At early times and large system sizes,
$\log (2\pi Dt)\ll \log\frac{N_0}{\rho_\ell}\propto \log L$, so that
$k$ is a function of $t/\log L$.  Note that assuming a rate of
reversals controlled by the slope of the front would lead to scaling
forms $P(\tau/\log L)$ and $\lambda(t/\log L)$ during the early
Gaussian spreading of the flock.

After a time $t_{\rm prot}$ that can be estimated as the inflexion
point in Fig.~\ref{fig:lambda}a, the protrusion grows out of the
main peak. Afterwards, the diffusive spreading of the main peak does
not affect the leading edge anymore.  At this stage, $\lambda(t)$
keeps increasing for some time, since the flipping of larger
protrusions is more likely to generate a peak with a mass sufficient
to revert the whole flock. In addition, $\lambda(t)$ also increases
after $t_{\rm prot}$ because of the slight dispersal in flock shapes
shown in Fig.~\ref{fig:shape}: as the time since the last reversal
increases, so does the fraction of flocks with a liquid protrusion.
For $t>t_{\rm prot}$, the $\tau/\log L$ scaling stops since the
leading edge converges to a well-defined steady profile. The time
$t_{\rm prot}$ can then be estimated by the time it takes for the
front to reach the slope corresponding to the liquid protrusion, so
that $t_{\rm prot}\propto \log L$.

We thus expect two different scaling regimes: an initial Gaussian
spreading leading to $\lambda(t/\log L)$ 
and a late-time scaling form $\lambda(t-\alpha \log L)$, once the
protrusion has grown out of the main peak. As shown in
Fig.~\ref{fig:lambda}c-d, the protrusion grows out quite early so
that the first regime is never quantitatively observed in our
simulations: most reversals take place after the protrusion has grown
out. This explains why
the $\lambda(t-\alpha \log L)$ and $P(\tau-\alpha \log L)$ scalings work satisfactorily.

\subsection{Role of the parameters $D$, $\eps$ and $\beta$}
\label{sec:flocks-v}

\begin{figure}
    \centering
    \includegraphics[width=0.98\columnwidth]{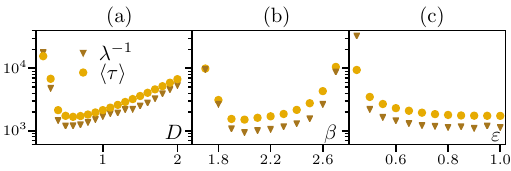}
    \caption{Mean reversal time $\langle \tau\rangle$ as a function of
      $D$ for $\beta=1.8$ and $\varepsilon=0.8$ {\bf (a)}, of $\beta$
      with $D=0.5$ and $\varepsilon=0.8$ {\bf (b)}, and of $\varepsilon$
      with $\beta =2 $ and $D=0.5$ {\bf (c)}. The data are compared
      with the characteristic time $\lambda^{-1}$ that corresponds to
      a purely exponential distribution
      $P_{\rm exp}(\tau)=\lambda e^{-\lambda \tau}$, with $\lambda$
      measured from the tail of the true distribution
      $P(\tau)$. Parameters: $\rho_0=3$, $L=1000$.}
    \label{fig:tau-v}
\end{figure}

Most results presented so far were obtained at fixed, rather typical,
parameter values.  We now report on the effect of changing the three
parameters $D$, $\eps$ and $\beta$.  Figure~\ref{fig:tau-v} shows how
$\langle \tau \rangle$ and $\lambda$, the reversal rate extracted from
the exponential tail of $P(\tau)$, vary with each of these parameters,
keeping the other two constant.

Comparing $\langle \tau\rangle$ with $\lambda^{-1}$ (which is the mean
of a normalized exponential distribution with rate $\lambda$), we see
that $\lambda^{-1}$ `underestimates' $\langle \tau\rangle$ but that
both quantities essentially vary in the same manner.  (Their
difference is due to the transient regime before $\lambda(t)$ reaches
its asymptotic value, as shown in Fig.~\ref{fig:lambda}a-b.)
To account for the variations of $\langle \tau \rangle$, we can thus
focus on the reversals at late times, which take place after the
protrusion has developed at the front of the flocks.

\begin{figure}
    \centering
    \includegraphics[width=0.48\columnwidth]{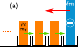}
    \vspace{0.3cm}
    \includegraphics[width=0.48\columnwidth]{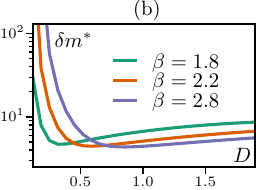}
    \includegraphics[width=0.48\columnwidth]{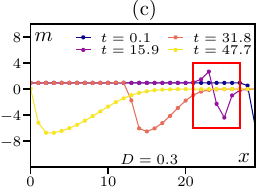}
    \includegraphics[width=0.48\columnwidth]{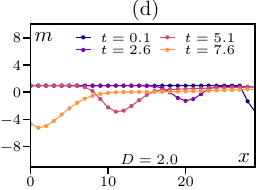}\\
    \includegraphics[width=0.48\columnwidth]{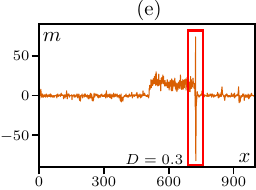}
    \includegraphics[width=0.48\columnwidth]{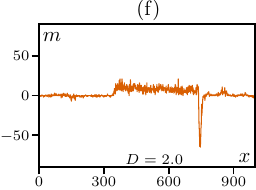}
    \caption{{\bf (a):} Sketch of the initial condition in the reduced
      mean-field model. {\bf (b):} Minimal fluctuation $\delta m^*$
      necessary to reverse the flock in the reduced model. {\bf
        (c,d):} Evolution of the density profile in the reduced model
      at low $D=0.3$ (c) and high $D=2.0$ (d) for fluctuations
      $\delta m=8$ (c) and $\delta m=6$ (d), slightly bigger than
      $\delta m^*$. Other parameters: $\eps=1$ and $\beta=2$. {\bf (e,f)} Instantaneous
      profiles in the microscopic model at the beginning of a reversal
      at low $D=0.3$ (e) and high $D=2$ (f). The transient aster (red
      boxes) observed at low $D$ in the reduced model also appears in
      the full AIM. $\rho_0=5$, $L=1000$, $\eps=1$ and $\beta=2$.}
    \label{fig:reduced}
\end{figure}

We find that the divergences of $\langle \tau \rangle$ visible on Fig.~\ref{fig:tau-v}
can be explained by two different mechanisms. First, decreasing $D$ or $\eps$
or increasing $\beta$ while keeping the other parameters fixed brings the
system from the flocking phase to the aster phase (see the phase diagrams
in Fig.~\ref{fig:phase-diagrams}). Close to the transition, 
transient asters appear at the beginning of an attempted
reversal, as shown in Fig.~\ref{fig:reduced}e. Once the aster is formed, the
propagation of the fluctuation stops until the aster dissolves. (See section Sec.~\ref{sec:asters} for a detailed discussion on aster dynamics.) In the mean time, a large amount of particles arrive from the flock, which tends
to destroy the aster and resume the forward motion of the flock. This process thus tends to suppress
reversals: transient asters
protect the flock against fluctuations, hence increasing their lifetime. 

The other divergences of $\langle \tau\rangle$, in the large $D$ and small $\beta$ limits, can be accounted for by comparing the roles of diffusion and alignment during a reversal. Aligning interactions tend to flip particles from the flock, hence strengthening the fluctuation. On the contrary, diffusion damps the fluctuation as it propagates. Diffusion dominates when $D/e^\beta$
---the ratio of hopping to alignment rates--- increases, i.e. when $D$ increases or $\beta$ decreases. 
Figure~\ref{fig:tau-v} shows a steeper divergence when varying
$\beta$ compared to varying $D$, as expected from this reasoning.

To support these two scenarios, we now evaluate how likely a
fluctuation is to reverse a flock in an idealized situation. We
consider the deterministic mean-field evolution of an initial
fluctuation of tunable size that encounters an ordered phase mimicking
the protrusion of a flock at density $\rho_\ell$ and magnetization
$m_\ell$ such that $m_\ell/\rho_\ell=\tanh(\beta m_\ell/\rho_\ell)$
(its mean-field value). 

The initial condition is depicted in
Fig.~\ref{fig:reduced}a and the evolution is that of
Eqs.~(\ref{eq:hydro-rho})-(\ref{eq:hydro-m}) after one takes a
mean-field approximation for the nonlinear terms. We choose
$\rho_\ell=1$ without loss of generality, thus fixing the unit of
density, and take the initial fluctuation to be fully ordered with
magnetization $-\delta m$ and density $\delta m$, propagating to the
left. We observe that there is a value $\delta m^*$ such that small
fluctuations $\delta m<\delta m^*$ do not propagate and the system
remains ordered, whereas large fluctuations $\delta m>\delta m^*$
propagate and flip the entire initial flock~\footnote{In practice we
  choose as a criterion that the fluctuation has propagated if the
  magnetization at a distance $L=30$ sites from the initial
  fluctuation has changed sign.}. This critical fluctuation size
varies with the parameters, as shown in Fig.~\ref{fig:reduced}b where we
vary $D$ for several values of $\beta$. At small $D$ we see the
transient asters observed in the microscopic model whereas they are
not observed for larger $D$ (see Fig.~\ref{fig:reduced}c-d). The
variations of $\delta m^*$ are consistent with the variations of
$\langle \tau \rangle$ in the microscopic model as we see a sharp
increase at small $D$ corresponding to the appearance of transient
asters and a slower increase at high $D$ with no asters. This minimal model of reversals, despite its simplicity,  thus reproduces the basic features of the mean
reversal time and supports the two  scenarios outlined above
for the divergences of $\langle \tau \rangle$.

\subsection{Coarsening}
\label{sec:flocks-coarsening}

\begin{figure}
  \centering
      \includegraphics[width=0.83\columnwidth]{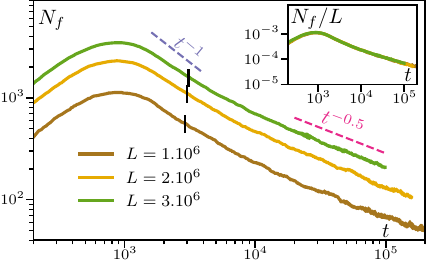}
      \caption{Number of flocks $N_f$ as a function of time, starting
        from a homogeneous disordered initial conditions. The short
        vertical black lines indicate the crossover time $t_c$ as
        given in Eq.~(\ref{eq:tcross}). $\rho_0=10$, $\beta=2$,
        $D=0.5$, $\eps=1$.}
    \label{fig:coarsening}
\end{figure}

So far we only considered the steady-state regime where the system
contains a single flock. We now study the coarsening dynamics that
bring the system from a disordered initial condition to such a steady state.
As shown on the space-time diagram of
Fig.~\ref{fig:phases}e, many small flocks form at early times and
merge when they encounter until only one remains. 
The evolution of the number of flocks $N_f$ can be written as
\begin{equation}
  \label{eq:Nf-coarsening}
  \dot N_f(t) =-\frac{N_f(t)}{t_{\rm coll}(t)}
\end{equation}
with a function $t_{\rm coll}(t)$ that may depend on $N_f$ and which
we can interpret as the mean time before a flock collisions with a
neighboring one. 

We can anticipate two regimes for this collision time.
At early times, when flocks are close to one another, they typically
  encounter a neighbouring flock before reversing their direction. The
  collision time will then be the ballistic time
  $t_{\rm coll}^B=\ell/v$ with $\ell=L/N_f$ the mean distance between
  flocks.
At later times, flocks are further apart and thus reverse their
  direction of motion before colliding. The collision time then has a
  diffusive scaling $t_{\rm coll}^D=\ell^2/(2D_{\rm eff})$ with
  the effective diffusion coefficient $D_{\rm eff}=v^2\langle\tau\rangle$.
The crossover between these two regimes is expected at a time $t_c$
such that $t_{\rm coll}^B(t_c)=\langle \tau \rangle$. Solving
Eq.~(\ref{eq:Nf-coarsening}) in the ballistic regime gives
\begin{equation}
  \label{eq:Nf-ballistic}
  N_f(t)=\frac{L}{\frac{L}{N_f(0)}+v t}
\end{equation}
from which we deduce that the crossover time is 
\begin{equation}
  \label{eq:tcross}
  t_c=\langle \tau\rangle-\frac{L}{v N_f(0)}.
\end{equation}
Neglecting the dependence of $\langle \tau\rangle$ on $\log L$, which
would give subdominant corrections, the number of flocks in the
diffusive regime then follows $\dot N_f\propto -N_f^3$ and thus
$N_f(t)\sim 1/\sqrt{t}$.

Comparing with simulations in Fig.~\ref{fig:coarsening}, we see that
the late-time coarsening is indeed clearly diffusive. At short time, we first see an increase in the number of flocks corresponding to the time that they form and grow large enough to be detected by our algorithm (we use a system of two thresholds at $\rho=2$ and $\rho=8$ with spatial smoothing on the length $\delta x=50$ to detect the flocks robustly). The crossover time $t_c$ is then computed from Eq.~(\ref{eq:tcross}), taking the initial time to be the time with the maximum number of flocks, and
we indicate $t_c$ by black lines in Fig.~\ref{fig:coarsening}. Consistently with the analysis above, we do see a faster-than-diffusive coarsening in the short window before $t_c$.

\section{Asters}
\label{sec:asters}

At low temperature or small velocity, 
a phase develops in which static
structures coexist with a dilute gas. (See the phase diagrams in
Fig.~\ref{fig:phase-diagrams} and the snapshot in
Fig.~\ref{fig:phases}d.) Each of these `asters' consists in two peaks of right
and left-moving particles which apparently `block' each
other. Remember, however, that there is no exclusion in our model so
that the underlying mechanism is necessarily more complex and calls
for an explanation. We termed these structures asters, by analogy to
the star-shaped defects observed in 2D active
systems~\cite{kruse2004asters,farrell2012pattern}.
\begin{figure}
    \centering
    \includegraphics[width=0.95\columnwidth]{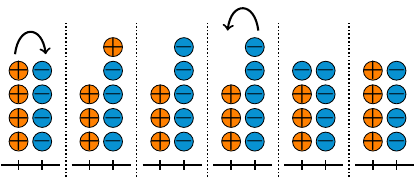}

    \caption{A typical sequence of configurations that make asters
      stable at low temperature.}
    \label{fig:aster_scenario}
\end{figure}

A closer look at the microscopic dynamics of an aster reveals that its
stability arises from periodic orbits in configuration space, as
illustrated in Fig.~\ref{fig:aster_scenario}. Starting from two peaks
of opposite magnetization, one particle---say with positive
spin---hops forward and lands onto the second site where it belongs to
the minority phase. It then flips and aligns with its new
environment. The most likely move is then that one of the minus
particles hops forward onto the site populated by plus particles. Once
again, this particle belongs to the minority phase and flips, leading
the system back into its original state. At low temperatures, such
trajectories are much more likely than trajectories leading to the
evaporation of the aster. To quantify the stability of asters, we
first consider in Sec.~\ref{sec:asterT0} a zero-temperature
fully-asymmetric model in which the lifetime of an aster can be
computed exactly. We then discuss the stability of asters at finite
temperature and the corresponding coarsening dynamics in
Sec.~\ref{sec:asterTsmall}. Finally, we show in
Sec.~\ref{sec:astersProfile} that, for $0<\eps<1$, asters typically
have a richer shape than suggested by Fig.~\ref{fig:aster_scenario}:
they are not perfectly localized and admit an exponential tail with a
characteristic size. Using a mean-field approximation, we compute the
corresponding decay length and determine a necessary condition for the
existence of asters that captures qualitatively, albeit not
quantitatively, the phase boundary between flocks and asters shown in
Fig~\ref{fig:phase-diagrams}.

\subsection{Lifetime of an aster at zero temperature}
  \label{sec:asterT0}

  To make progress analytically, we first consider a $T=0$, fully
  asymmetric version of the AIM. Particles hop at rate $v=2D$ and the
  aligning interaction is resolved instantaneously, due to the
  $\beta\to\infty$ limit of the flipping
  rate~\eqref{eq:flipping-rates}.  In practice, when a particle hops
  onto a new site with two particles or more, it immediately acquires
  the magnetization of the target site. The
  limit is ill defined when one has both $m_i=0$ and
  $\beta=\infty$. For simplicity, we here assume that when a particle
  of spin $s$ arrives on a site occupied by a particle of spin $-s$,
  it flips and acquires a $-s$ spin. As a result, the sole microscopic
  time-scale of the system is $v^{-1}$. The case in which both
  particles keep their current spins and flip at rate $\omega_0$ is
  qualitatively similar ---albeit more involved due to the presence of
  the second time scale $\omega_0^{-1}$--- as discussed in
  Appendix~\ref{Sec:0Talpha}.

\begin{figure}
  \includegraphics[width=\columnwidth]{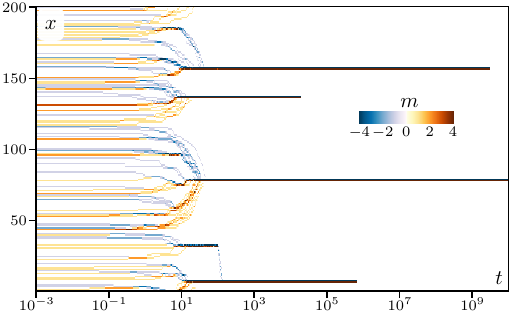}
  \caption{Starting from a random configuration at $t=0$ in the
    zero-temperature model, the system rapidly evolves into a large
    number of small asters. These coarsen on much longer time
    scales, as smaller condensates evaporate and are redistributed into
    larger ones. Parameters: $L=200$, $N=100$, $v=1$, $\eps=1$.}
  \label{fig:dynamics0T}
\end{figure}

Consider an initial condition in which $N$ particles are randomly
placed on the lattice and each site is given a magnetization at
random. The dynamics is such that all sites remain fully ordered at
all times, with $m_i=\pm \rho_i$, and the total escape rate is always
$N v$. A typical trajectory is shown in Fig.~\ref{fig:dynamics0T}.  It
rapidly leads to the emergence of an extensive number of asters, whose
rare and sudden evaporations lead to a coarsening dynamics. To
characterize the latter, we compute the lifetime of an aster
comprising $\na$ particles (spread out over the two sites).

\begin{figure}
  \includegraphics[width=.95\columnwidth]{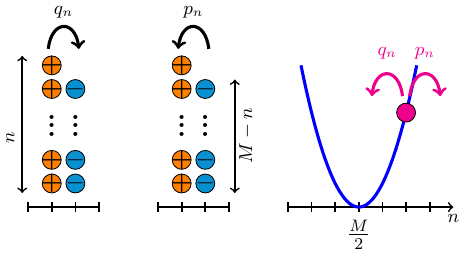}
  \caption{{\bf Left:} A particle hops out of a site with $n$
    particles at a rate $q_n=n v$. This takes the aster from
    $(n,\na-n)$ to $(n-1,\na-n+1)$. {\bf Center:} A particle hops out of
    a site with $\na-n$ particles at a rate $p_n=(\na-n) v$. This
    takes the aster from $(n,\na-n)$ to $(n+1,\na-n-1)$. {\bf Right:}
    This dynamics is equivalent to the Brownian dynamics of a single
    particle at position $n$, confined in a harmonic
    potential $V(n)=\frac v 2 (\frac{n}{M}-\frac{1}2)^2$, with $n=0$ and
    $n=\na$ being absorbing boundaries corresponding to the
    evaporation of the aster. }
  \label{fig:singleaster}
\end{figure}

Let us consider the situation depicted in Fig.~\ref{fig:singleaster}
in which an aster is formed with $n$ particles with $+1$ spins at site
$i$ and $\na-n$ particles with $-1$ spins at site $i+1$. We denote
this configuration as $\bfn\equiv(n,\na-n)$. The system evolves at
rate $q_n=v n$ towards the configuration ${\bf n-1}$ and
$p_n=v(\na-n)$ towards ${\bf n+1}$. We denote by $r_n=p_n+q_n$ the
escape rate from configuration $\bfn$, which is here given by $r_n=v
\na$. Given the expressions of $q_n$ and $p_n$, there is a linear
drift that takes the system towards the most likely configuration
$\bfn={\bf \na}/2$. Note that the configurations ${\bf \na}$ and ${\bf
  0}$ are limiting cases that correspond to the evaporation of the
aster. To get an intuitive understanding of the aster dynamics, we
consider the master equation of this process:
\begin{equation}\label{eq:MEPofN}
  \partial_t P(n)=q_{n+1} P(n+1)+p_{n-1}P(n-1)-(q_n+p_n) P(n)\;.
\end{equation}
Small Gaussian fluctuations close to the most likely configuration
$\bfn={\bf \na}/2$ are well described by introducing $P(x=n/\na)=\na P(n)$ and
expanding~\eqref{eq:MEPofN} to first order in $dx=\na^{-1}$. Doing so
yields the Fokker-Planck equation
\begin{equation}\label{eq:MEPofx}
  \partial_t P(x)=\frac{\partial}{\partial x}\left [ V'(x) P(x) + \frac{v}{2 \na} \frac{\partial}{\partial x} P(x) \right]\;,
\end{equation}
which corresponds to the dynamics of a Brownian particle in a harmonic
potential $V(x)=v(x-\frac{1}{2})^2$ at temperature $\tfrac{v}{2\na}$, as illustrated on
Fig.~\ref{fig:singleaster}. The mean-first passage time $\tau$ until
the evaporation of an aster with $\na$ particles can then be
estimated using the Arrhenius scaling $\log \tau \propto \na/2$. This
scaling is not expected to hold quantitatively, since the diffusive
approximation~\eqref{eq:MEPofN} is expected to fail in the
large-deviation regime where $x\gg \frac{1}{\sqrt \na}$, but it
captures the physics that makes the aster long-lived at zero
temperature.

We now wish to compute exactly the mean-first passage time to reach
either ${\bf \na}$ or ${\bf 0}$, which will trigger the redistribution
of the aster particles into a neighboring aster, hence driving the
coarsening process.
Let us note $T_n$ the average evaporation time, starting from
configuration $\bfn$. In an average time $r_n^{-1}$, the system jumps
to $\bfn-1$ or $\bfn+1$ with probabilities $q_n/r_n$ and $p_n/r_n$,
respectively. One thus has the recursive relation:
\begin{equation}\label{eq:recurs}
  T_{n}=\frac{1}{r_n} + \frac{p_n}{r_n}\,T_{n+1} + \frac{q_n}{r_n}\, T_{n-1}
\end{equation}
that has to be solved with the boundary conditions
\begin{equation}
  T_{0}=T_{\na}=0\;.
\end{equation}
To do so, we follow standard
methods~\cite{van1992stochastic,antal2006fixation} and introduce
$U_n=T_{n-1}-T_n$. Equation~\eqref{eq:recurs} can then be rewritten as
\begin{equation}\label{eq:recurs2}
  p_n U_{n+1}=q_n U_n+1\;.
\end{equation}
Introducing $\pi_0=1$ and, for $i\geq 0$,
\begin{equation}
  \pi_{i}\equiv \prod_{k=1}^i \frac{q_k}{p_k} \;\;\;\;{\rm and, further,}\;\;\;\;  \sigma_i = \sum_{k=1}^i \frac{1}{p_k\pi_k}\;,
\end{equation}
Eq.~\eqref{eq:recurs2} is readily solved as
\begin{equation}
  U_{n+1}=\pi_n U_1 + \pi_n \sigma_n\;.
\end{equation}
Using $T_0=0$, we first get $T_1=-U_1$. The definition of $U_n$ then
recursively leads to
\begin{equation}\label{eq:T_n}
  T_{n\geq 2}=-U_1 \sum_{i=0}^{n-1}\pi_i-\sum_{i=1}^{n-1}\left[\pi_i \sigma_i \right]\;.
\end{equation}
\if{Finally, $U_1$ is determined by imposing the boundary condition
$T_{\na}=0$ and using Eq.~\eqref{eq:recurs} \red{Eq.~\eqref{eq:T_n} gives the same result w/ easy computations or am I missing something ?}. Tedious but
straightforward algebra leads to}\fi
Finally, $U_1$ is determined by imposing the boundary condition $T_{\na}=0$, which gives:
\begin{equation}
  U_1=-\frac{\sum_{i=1}^{\na-1}\left[ \pi_i \sigma_i \right]}{\sum_{i=0}^{\na-1} \pi_i }
  \;.
\end{equation}
All in all, the mean time to evaporation starting from the
configuration $\bfn$ is given by
\begin{equation}\label{eq:Tn}
  T_{n}=\frac{\sum_{i=0}^{n-1}\pi_i}{\sum_{i=0}^{\na-1}\pi_i} 
  \sum_{i=1}^{\na-1}\left[\pi_i\sigma_i \right]-\sum_{i=1}^{n-1}\left[\pi_i\sigma_i\right]\;
\end{equation}
where the second sum vanishes for $n=1$. 
The above results are valid for generic random walks with non-vanishing rates $p_n$ and $q_n$. 
In the case at hand, using $p_n=v(\na-n)$ and $q_n=v n$, we find
\begin{equation}\label{eq:rhoi}
  \pi_i=\binom{\na-1}{i}^{-1}\;,
\end{equation}
which allows computing $T_n$.

\begin{figure}
  \includegraphics[width=.48\columnwidth]{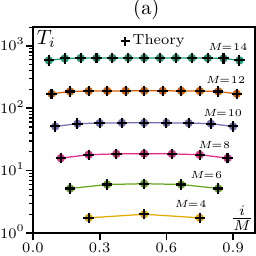}
  \includegraphics[width=.48\columnwidth]{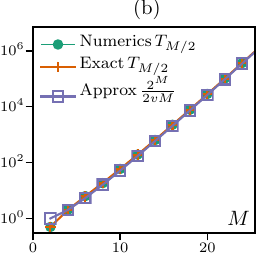}
  \caption{{\bf Left:} Mean-first-passage times from site $ i$ to sites
    $0$ or $\na$, as $\na$ varies from 4 to 14. The circles correspond
    to numerical simulations whereas the crosses correspond to
    Eq.~\eqref{eq:Tn}. {\bf Right:} Mean-first passage times to evaporation for an aster with $\na$ particles.  
    Green disks are results of numerical simulations, averaged over $10^4$ realizations. 
    Orange crosses correspond to the exact formula~\eqref{eq:Tn}, and purple squares to approximation~\eqref{eq:Tnapprox}.} 
    \label{fig:MFPTFromSites}
\end{figure}

The mean first passage times to evaporation starting from all possible
configurations are plotted in Fig.~\ref{fig:MFPTFromSites} (left), for $\na$ up
to 14. Interestingly, $T_n$ becomes rapidly independent of $n$. This
can be understood as follows: due to the asymmetry between $p_n$ and
$q_n$, the most likely path to evaporation is to fall from $n$ to $\na/2$
and then to get absorbed. Furthermore, Fig.~\ref{fig:MFPTFromSites} (left)
suggests an exponential increase of $T_n$ with $\na$. To compute the
leading order of $T_n$, consider the various terms of Eq.~\eqref{eq:Tn} and
their scaling as $\na\to \infty$. We first note, using
Eq.~\eqref{eq:rhoi}, that
  $\pi_0=\pi_{\na-1}=1$, $\pi_1=\pi_{\na-2}=(\na-1)^{-1}$, and  $\pi_{1< i<\na-2}={\cal O}({\na}^{-1})$.
  Consequently, $\pi_1+\pi_2+\dots+\pi_{n-1}\to 0$ for n<M and only
  $\pi_0$ and $\pi_{M-1}$ contribute to the sums.
\if{\begin{align}
  \sum_{i=0}^{\na-1}\pi_i &\simeq 2;\\
  \sum_{i=0}^{n-1}\pi_i &\simeq 1\;\text{for}\;n<\na,\\
  \sum_{i=1}^{n-1}\pi_i &\to 0\;\text{for}\;n<\na,\\
\end{align}}\fi
Noting also that $(\pi_k p_k)^{-1}$ simplifies into
\begin{equation}
\frac{1}{\pi_kp_k}=\frac{(\na-1)!}{v k! (\na-k)!}=\frac{1}{v \na} \binom{\na}{k}\;,
\end{equation}
we see that Eq.~\eqref{eq:Tn} can be approximated as
\begin{equation}\label{eq:Tn2}
  T_{n}\sim \frac{1}{2 v \na}\sum_{i=1}^{\na-1}\left[\pi_i\sum_{k=1}^i\binom{\na}{k}\right]-\frac{1}{v\na}\sum_{i=1}^{n-1}\left[\pi_i\sum_{k=1}^i\binom{\na}{k}\right]\;.
\end{equation}
Finally, we note that, for $n<\na-1$,
\begin{align}
\sum_{i=1}^{n}\left[\pi_i\sum_{k=1}^i\binom{\na}{k}\right]&<\left[\sum_{i=1}^{n} \pi_i\right]\left[\sum_{k=1}^{\na-2}\binom{\na}{k}\right]\\&=o\left(\sum_{k=1}^{\na-1}\binom{\na}{k}\right)\;,
\end{align}
so that
\begin{equation}\label{eq:Tnapprox}
  T_{n}\sim \frac{1}{2 v \na}\pi_{\na-1}\sum_{k=1}^{\na-1}\binom{\na}{k}\sim \frac{ 2^{\na}}{2 v \na}
\end{equation}
Figure ~\ref{fig:MFPTFromSites} (right) shows the comparison between numerical
measurements of $T_{\na/2}$, its exact expression~\eqref{eq:Tn} and
the asymptotic estimate~\eqref{eq:Tnapprox}, which is remarkably close
to the exact values. The stability of asters and their coarsening at
zero temperature thus stems from an evaporation rate that vanishes
exponentially with their density.

\subsection{Coarsening at small temperature}
\label{sec:asterTsmall}

At zero temperature, aster coarsening occurs via an extremal dynamics:
smallest asters evaporate first and are redistributed among their
neighbors. Since the lifetime of an aster diverges exponentially
with its height, this leads to an extremely slow dynamics, as apparent
from Fig.~\ref{fig:dynamics0T}.

\begin{figure*}
  \centering
    \includegraphics[width=.75\columnwidth]{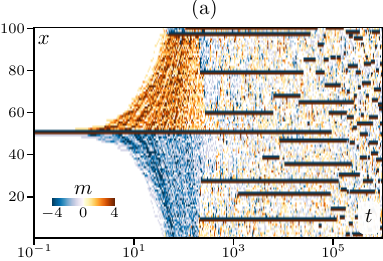}
    \includegraphics[width=.75\columnwidth]{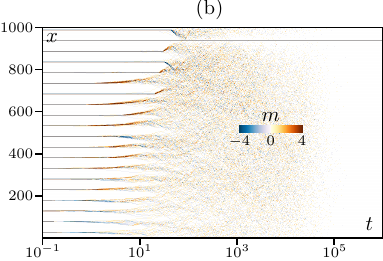}
    \includegraphics[width=.5\columnwidth]{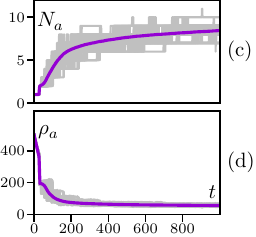}
     \caption{{\bf (a)} Spatiotemporal plot of the relaxation dynamics from a single aster to 
    	the fluctuating aster phase of the
    	AIM with rates~\eqref{eq:flipping-rates} 
    	($T=0.2$, $\eps=1$, $L=100$, $\rho_0=10$, $D=0.5$). 
    	{\bf (b):} same as (a) but for the
    	AIM with rates~\eqref{eq:unbounded}, starting from $20$ asters
    	whose initial density increases linearly from $6$ to $25$ as $x$
    	increases; coarsening leaves a single macroscopic
    	aster ($D=1$, $L=1000$, $\omega_0=1$, $\beta=0.375$,
    	$\rho_0=0.29$. 	{\bf (c-d):} Relaxation dynamics in the aster phase for the
    	AIM with rates~\eqref{eq:flipping-rates}. Number of asters as a function of time, averaged over
    	$1000$ simulations (magenta) (c) and average density contained in
    	one aster (d). Fifty representative trajectories appear in
    	gray. Parameters as in (a).}
    	\label{fig:asters_coarsening}
 \end{figure*}

%
%
%
%

At finite temperature, a new phenomenon sets in: when a particle hops
forward, as in the second configuration of
Fig.~\ref{fig:aster_scenario}, it now has a finite probability $P_{\rm
  hop}$ of hopping further forward and leaving the aster before flipping its spin, with 
\begin{equation}
  P_{\rm hop}=\frac{p}{p+W(s\to -s)}\;.
\end{equation}
In an aster comprising two sites with $\rho$ particles, this leads to
a flux of particle $  j_{\rm leak}(\rho)$ leaving the aster, where
\begin{equation}
  j_{\rm leak}(\rho) \propto p \rho P_{\rm hop}= \frac{p^2}{p+W(s\to-s)}\rho\;.
\end{equation}
The variations of $j_{\rm leak}(\rho)$ with $\rho$ then determine the
late-stage dynamics. For the flipping rates~\eqref{eq:flipping-rates},
$j_{\rm leak}(\rho) \propto \rho$ at large densities since the hopping
rate is bounded, $W(s\to-s)<\exp(\beta)$: large asters leak particles
faster than smaller ones, which arrests the coarsening and lead to a
steady-state with an extended number of finite-size asters, as
illustrated in Fig.~\ref{fig:asters_coarsening}a and
Fig.~\ref{fig:asters_coarsening}c-d. (It would be interesting to
 generalize approaches developed in the past to predict the size of
competing finite condensates~\cite{thompson2010zero} but this is
beyond the scope of this study.) For the unbounded rates studied
in~\cite{kourbane2018exact}:
\begin{equation}\label{eq:unbounded}
  W(s\to -s) = \omega_0 e^{-\beta s m_i},
\end{equation}
larger asters leak slower, leading to a single macroscopic aster, as
illustrated in Fig.~\eqref{fig:asters_coarsening}b. Finally, in a
related 1D flocking model, a coarsening into a single structure
reminiscent of an aster was observed in the presence of ``centering''
interactions, which bias the motion of particles towards dense
regions~\cite{raymond_flocking_2006}. The generality of the mechanisms leading
to aster-like structures and their stability is thus an interesting
question beyond the sole case of the AIM.

\subsection{Shape of asters}
\label{sec:astersProfile}
\begin{figure*}
    \centering
    \includegraphics[width=0.32\textwidth]{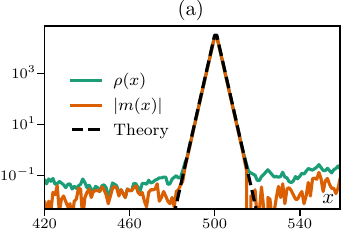}
    \includegraphics[width=0.32\textwidth]{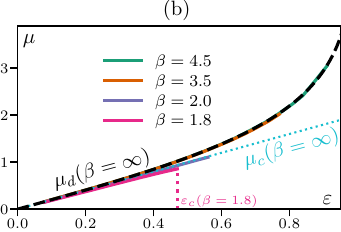}
    \includegraphics[width=0.32\textwidth]{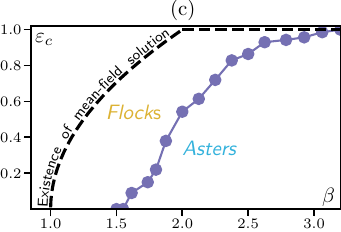}
    \caption{\textbf{(a):} Profile of an aster in the microscopic
      model. The dashed lines are the exponential decays predicted by
      Eqs.~(\ref{eq:sol-aster-phi-discret})-(\ref{eq:sol-aster-mu-discret}).
      $L=1000$, $\rho_0=100$, $\beta=4$, $\eps=0.5$,
      $D=0.5$. \textbf{(b):} Measured values of $\mu$ in the microscopic
      model at different $\beta$ compared with the predictions from the
      continuous mean-field equations (blue dotted line) and from the
      discrete ones (black dashed line). $L=1000$, $\rho_0=100$, $D=0.5$. \textbf{(c):} Phase boundary
      $\eps_c(\beta)$ such that asters are observed for $\eps<\eps_c$ in
      microscopic simulations (symbols), compared to the limit of
      existence of aster solutions in the mean-field equations (dashed
      line). $L=500$, $\rho_0=10$, $D=0.5$.}
    \label{fig:asters_theory}
\end{figure*}

Let us now consider the case $\eps<1$, where hopping is not fully
asymmetric. Thanks to diffusion, particles can hop backward and 
asters are then spread over several sites. Since they typically
comprise many particles, we expect to be able to describe their shape
using the mean-field equations (\ref{eq:MF2-rho},\ref{eq:MF2-m})
introduced in Sec.~\ref{sec:flocks-shape}.
It is straightforward to check that these equations
admit stationary solutions of the form 
\begin{equation}
  \label{eq:sol-aster}
  \rho(x)= k e^{-\mu_c x};\quad m(x)=k\phi e^{-\mu_c x}
\end{equation}
with $k$ a normalization constant and $\phi$ and $\mu_c$ (the
subscript ``c'' stands for ``continuous'' for a reason that will
become apparent shortly) solutions of
\begin{align}
  \label{eq:sol-aster-phi}
 &\frac{v^2}{2D} \phi(\phi^2-1)=\phi \cosh(\beta\phi)-\sinh(\beta\phi)\\
  \label{eq:sol-aster-muc}
  &\mu_c=-v\phi/D.
\end{align}
Furthermore, these are the only stationary
solutions of Eqs.~(\ref{eq:MF2-rho},\ref{eq:MF2-m}) such that
$\rho(x)$ and $m(x)$ are proportional to each other. Note that
Eq.~\eqref{eq:sol-aster} does not describe the full aster but only
half of it, for $x>0$ or $x<0$. We are not aware of any analytic
solution describing the full aster, including its singularity
at $x=0$. Finally, the competition between the asymmetric hops leading
particles towards the aster and the diffusive dynamics allowing them
to explore neighboring sites is reminiscent of the diffusive motion of
colloids under gravity. The solutions~\eqref{eq:sol-aster} can thus be
seen as two exponential atmospheres pointing towards the core of the
aster.

We now compare these solutions with the aster profiles measured in the
microscopic model in Fig.~\ref{fig:asters_theory}a. The profiles of
density and magnetization indeed exhibit an exponential decay on both
sides of the aster. In Fig.~\ref{fig:asters_theory}b, we see that the
measured decay exponent $\mu$ agrees well with the solution $\mu_c$ of
Eq.~(\ref{eq:sol-aster-muc}) at small $\eps$ but deviates for larger
$\eps$. Note that we plot only the curve $\mu_c(\beta=\infty)$ in
Fig.~\ref{fig:asters_theory}b since the curves for different values of
$\beta$ would be indistinguishable on the scale of the figure. Setting
$\beta=\infty$ also simplifies the calculation since it amounts to fixing
$\phi=\pm 1$ in Eq.~(\ref{eq:sol-aster-phi}) so that
$\mu_c(\beta=\infty)=\pm v/D=\pm 2\eps$.

Increasing $\eps$, we see that the discrepancy between $\mu_c$ and
the measured $\mu$ increases. This deviation can be attributed to the
continuous limit used to derive
Eqs.~(\ref{eq:MF2-rho},\ref{eq:MF2-m}).
If instead we retain the full
discrete equations Eqs.~(\ref{eq:hydro-rho},\ref{eq:hydro-m}) in
the mean-field approximation, and look in the same way for exponential
solutions
\begin{equation}
  \label{eq:sol-aster-discrete}
  \rho_i= k \kappa^i;\quad m_i=k\phi \kappa^i,
\end{equation}
we find the conditions
\begin{align}
  \label{eq:sol-aster-phi-discret}
  &\frac{4D(\kappa-1)^2\phi-v(\kappa^2-1)}{4\kappa}=\phi \cosh[\beta\phi]-\sinh[\beta\phi]\\
  \label{eq:sol-aster-mu-discret}
  &(\kappa+\frac{1}{\kappa}-2)=\frac{v\phi}{2 D}(\kappa-\frac{1}{\kappa});\quad \mu_d=-\log\kappa
\end{align}
The decay exponent $\mu_d$ (where ``$d$'' stands for
``discrete'') predicted by
Eqs.~(\ref{eq:sol-aster-phi-discret},\ref{eq:sol-aster-mu-discret})
is now in near-perfect agreement with microscopic simulations as shown
in Fig.~\ref{fig:asters_theory}b.

Interestingly, we find that there is a maximum value of $\eps$ that we
denote $\eps_c$ above which no physical exponential solution is found
(we require that $\phi\in[-1,1]$ and that $\mu_d$ be of opposite sign
to $\phi$ for a physical solution). In Fig.~\ref{fig:asters_theory}c,
we compare this limit value with the transition line between the
flock and aster phases in the microscopic model. Both lines show the
same trend as a function of $\beta$ but are quantitatively
different. This indicates that, unsurprisingly, exponential solutions
can exist at the mean-field level while asters are not observed
because either the full aster solutions do not exist at the mean-field
level or because they are unstable to fluctuations.

\section{Conclusion}

We have provided a detailed study of the active Ising model in one
space dimension. Despite its simplicity, the AIM shows two non-trivial
phases characterized by flocks and asters.  We found the flocking
phase to exhibit the same properties as in other one-dimensional
flocking
models~\cite{czirok_collective_1999,oloan_alternating_1999,raymond_flocking_2006,dossetti_cohesive_2011,solon_revisiting_2013,laighleis_minimal_2018,sakaguchi_flip_2019}:
Large ordered aggregates spread while propagating, merge when they
collide, and regroup when stochastically reversing their direction of
motion. As reported
before~\cite{oloan_alternating_1999,raymond_flocking_2006,solon_revisiting_2013},
the time between two reversals of a flock increases logarithmically
with system size. We went further and explained this dependence as
well as the variations with the velocity of particles and the
temperature. In addition, we have also analyzed in detail the flock
shapes and how they evolve in time. The global picture is that of a
liquid-gas phase separation which cannot relax to steady state because
of reversal events. Since this is expected to be generic for 1D
flocking models, we thus also expect our analysis of the reversal
times based on the flock shapes to be relevant to other systems.

We reported a new phase of the AIM. It is populated by ``asters''
which are composed of two peaks of opposite magnetization facing each
other. These asters appear here only because of the interplay of
alignment and self-propulsion and are thus qualitatively different
from traffic jams due to steric interactions or from the ``dipoles''
observed in Ref.~\cite{raymond_flocking_2006} which are produced by
centering interactions that favor motion towards high-density regions.
We have provided an exact solution of the zero-temperature dynamics of
an aster and showed that they dissolve in a time that is exponential
in their size. At infinite time, this leads to a single aster in the
zero-temperature limit but to a finite density of asters at small but
finite temperature.  Finally, we have shown that asters generically exhibit
exponential tails with decay exponents that are well predicted by a
mean-field theory.

Common statistical mechanics wisdom states that fluctuations become
more important as the dimension of space becomes lower. In
equilibrium, this goes as far as preventing the existence of an
ordered state in a model with only short-ranged interactions. In our
active system, the flocking phase shows that ordering is possible but
fluctuations still prevent true symmetry breaking by inducing
reversals of the flock direction. Fluctuations play a somewhat
different but also essential role in the dynamics of asters: we found
that the dissolution of asters and thus their coarsening is due to
rare fluctuations, giving rise to extremal dynamics. In contrast, we
found that some features of flocks and asters are accounted for to a
good accuracy by a mean-field approach that neglects fluctuations.

\begin{acknowledgements}
  We thank Mourtaza Kourbane-Houssene for his early involvement in the
  project. JT acknowledges financial support from ANR grant THEMA.  
\end{acknowledgements}

\appendix{}

\section{Numerical simulations}
\label{sec:numerics}

\subsection{Discrete-time simulations}

Our discrete-time simulations are implemented using random sequential
updates: Particles are chosen at random and updated sequentially. For
each update, the time is increased by $dt/N$. Denoting by $i$ the site
of the chosen particle and by $s$ its spin, the particle either hops,
flips its spin, or does nothing with probabilities:
\begin{equation}\label{eq:DTrates}
  \begin{cases}
    \text{Proba(hop to site $i+1$):}& D(1+s \eps) dt\\
    \text{Proba(hop to site $i-1$):}& D(1-s \eps) dt\\
    \text{Proba(spin flips):} & \omega_0e^{-\beta s \frac{m_i}{\rho_i}}dt\\
    \text{Proba(no update):} & 1-2D dt-\omega_0e^{-\beta s \frac{m_i}{\rho_i}}dt
  \end{cases}
\end{equation}
In the last case, we say that a move has been ``rejected''.

Let us note $\C_i$ the state of particle $i$, defined by its spin and
position. A configuration $\C$ is then given by $\C=\{\C_i\}$.
Consider the discrete-time Markov process defined by the above
algorithm. Since there is a $1/N$ probability of choosing a given
particle, the corresponding discrete-time master equation reads
\begin{eqnarray}
\notag
P\left(\C,t+\frac{dt}{N}\right) = \frac{1}{N} \left[\sum_i \sum_{\C_i'\neq\C_i} P(\C',t) W(\C'\to\C) dt \;\;\;\right.\\
 +  P(\C,t) \left.\sum_i  \left(1- \sum_{\C'_i\neq \C_i}  W(\C\to\C') dt\right)\right]\;\;\;\;\;\;\;\label{eq:DTME}
\end{eqnarray}
where $\C'$ is identical to $\C$ but with particle $i$ in state
$\C_i'$, and the rates $W$ are given by $W=D(1\pm \eps)$ or by
Eq.~\eqref{eq:flipping-rates}, depending on the transition. It is then
straightforward to check that the limit of~\eqref{eq:DTME} as
$N\to\infty$ is the continuous-time master equation of the AIM:
\begin{equation*}\label{eq:MECT}
  \partial_t P(\C,t)=  \sum_{\C_i'\neq\C_i} [P(\C',t)  W(\C'\to\C) - P(\C,t)  W(\C\to\C')]
\end{equation*}

This algorithm has two main limitations. First, it is only exact in
the large $N$ limit and it is thus not adapted to study small
systems. This precludes using it for the study of the finite-size asters
of section~\ref{sec:asterT0}. Then, its efficiency relies on the fact
that there are as few proposed updates as possible that lead to
rejection. Since the only constraint on the algorithm is that the
probabilities in~\eqref{eq:DTrates} are between 0 and $1$, we take the
largest possible $dt$, given by $dt=(2D+\omega_0e^\beta)^{-1}$, to
minimize the probability of a particle doing nothing during an update.

\subsection{Continuous-time simulations}
Our continuous time simulations were implemented as follows. For each
particle $i$, with spin $s_i$, we define an escape rate $r_i=
2D+\omega_i$, where $\omega_i$ is the rate at which $s_i$ flips into
$-s_i$. At time $0$, we sample a time $\tau_i$ for each particle
according to $P(\tau_i)=r_i e^{-r_i\tau_i}$. The algorithm then goes
iteratively over the following loop:
\begin{enumerate}
\item Set the simulation time to $t=\tau_j$ where $\tau_j=\underset{i}{\text{inf}}\,
  \tau_i $.
\item Update particle $j$ using tower sampling: pull a random number
  $\eta\in[0,r_j]$. If $\eta<D(1-s_j \eps)$, the particle hops
  to the left. If $D(1-s_j \eps)<\eta<2D$, the particle hops to
  the right. Otherwise, the spin flips: $s_j\to-s_j$.
\item Update any $r_k$ that has changed because of step 2 and sample
  new $\tau_k$'s according to $P(\tau_k)=r_ke^{-r_k(\tau_k-t)}\Theta(\tau_k-t)$. A new
  $\tau_j$ is also sampled, whether or not $r_j$ has changed.
\end{enumerate}
To make step 1 as efficient as possible, we use a heap structure,
which is a binary sorting tree consisting in a root and two sub
heaps. The heap is such that the time stored at a root is always
smaller that the times stored at the roots of the two sub heaps. Finding the
smallest time then has a cost of $\mathcal{O}(1)$ whereas the
reorganization of the heap due to step 3 has a cost
$\mathcal{O}(\log_2 N)$.

As we now discuss, the above algorithm, which is a generalization of
the celebrated Gillespie algorithm, is an exact implementation of the
continuous time Markov chain that defines the AIM. To show that an
algorithm simulates a Markov chain defined by transition rates
$W(\C\to\C')$, it suffices to show that trajectories are generated
according to two principles. First, the system transitions out of a
configuration $\C$ after a time $\tau$ distributed as
\begin{equation}\label{Poftau}
 \!\! P(\tau)\!=\! r(\C) e^{-\tau r(\C)}\;\;\text{where}\;\; r(\C)\!=\!\sum_{\C'\neq\C}\! W(\C\to\C')\;.
\end{equation}
Then, when a configuration change happens, the target configuration
$\C'$ is chosen with probability
\begin{equation}\label{Poftrans}
  P(\C\to\C')=\frac{W(\C\to\C')}{\sum_{\C''}W(\C\to\C'')}\,.
\end{equation}

Consider a case with $N$ possible transitions, with corresponding rates $W_i$, and
take $N$ transition times sampled according to
$p(\tau_i)=W_ie^{-W_i \tau_i}$. The probability that the smallest time
is larger than $\tau_{\rm min}$ is
\begin{equation*}
P(\underset{i}\inf\, \tau_i>\tau_{\rm \min})=\prod_i \int_{\tau_{\rm \min}}^\infty d\tau_i W_i e^{-W_i \tau_i}=e^{-\tau_{\rm \min}\sum_i W_i}
\end{equation*}
so that the probability density of the smallest time $\tau_{\rm \min}=\inf \tau_i$ is given by
\begin{equation*}
P(\tau_{\rm \min})=-\partial_{\tau_{\rm \min}} P(\underset{i}\inf
\tau_i>\tau_{\rm \min})=r e^{-\tau_{\rm \min} r}\;,
\end{equation*}
with $r=\sum_i W_i$, which indeed yields Eq.~\eqref{Poftau}. Tower
sampling, as done in step 2, is then a simple way to choose a
transition according to~\eqref{Poftrans}.

Note that our algorithm 1-3 relies on two more ingredients. First, for
$N$ particles, there are $3N$ possible moves, but we only use $N$
times $\tau_i$, grouping hops and flips to get an escape rate per
particle. This is largely arbitrary---and simply appropriate given our
data structure---and also leads to the exact implementation of the
continuous time Markov chain because the escape rates are linear
combinations of the possible rates. Further splitting the transitions
into sub-transitions leaves $P(\tau_{\rm min})$ unchanged since
$\sum_i W_i$ remains equal to the total escape rate. Second, we only
need to sample new times $\tau_k$ at step 3 if $r_k$ has changed, or
if the particle $k$ has just moved. This is because the dynamics is
Markovian so that any particle $n$ whose time $\tau_n$ is larger than
$t$ has a waiting time that is already correctly distributed. Indeed,
its next configuration change was sampled at an earlier time $\tau_n'$
with a law $P(\tau_n)=r_n e^{-r_n (\tau_n-\tau_n')}$. We now know that
$\tau_n$ is larger than the current time $t$, which had a probability
$P(\tau_n>t)= e^{-r_n(t-\tau_n')}$ to happen. Given that $\tau_n$ is
larger than $t$, $\tau_n$ is thus distributed according to the
conditional probability
$P(\tau_n)/P(\tau_n>t)=r_n e^{-r_n (\tau_n-t)}$, as it should at time
$t$.

\begin{figure}
\includegraphics[width=0.8\columnwidth]{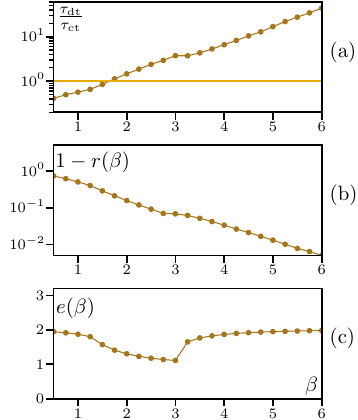}
\caption{Comparison between discrete- and continuous-time
  algorithms. {\bf (a):} Ratio between the cpu time used for a
  discrete-time simulation, $\tau_{\rm dt}$, and a continuous-time
  simulations, $\tau_{\rm ct}$, as $\beta$ is varied. The discrete-time simulation is faster
  at large temperature but its efficiency decreases exponentially with
  $\beta$. {\bf (b):} The rejection rate of the discrete-time
  algorithm, $r(\beta)$, converges exponentially to $r(\beta)=1$ as
  $\beta$ increases. {\bf (c):} The average escape rate per particle,
  $e(\beta)$, shows little variation as $\beta$ is varied and remains
  close to $r\simeq 2 D+\omega_0$. The dip for intermediate $\beta$
  corresponds to the flocking phase where most particles are
  aligned. Parameters: $D=0.5$, $\eps=1$, $\omega_0=1$, $L=500$,
  $\rho_0=10$. In panel (a), the system is simulated up to a final
  time $t=110\,000$. }\label{fig:simulation times}
\end{figure}

The clear advantages of this continuous-time algorithm are that it is
exact and that there is no rejection. At every move, however, several
new times $\tau_k$ have to be sampled, the heap has to be reorganized,
and the data structure is much heavier than for the discrete-time
algorithm. As a result, the discrete-time algorithm is always more
efficient when the rejection rate is low. For our simulations, the
discrete-time simulations were faster at high temperatures, whereas
the continuous-time simulations were faster at lower temperature (see
Fig.~\ref{fig:simulation times}) and we used them accordingly.

\subsection{Numerical procedure to spatially align flocks.}
\label{sec:numerics-align}
\begin{figure}
  \includegraphics{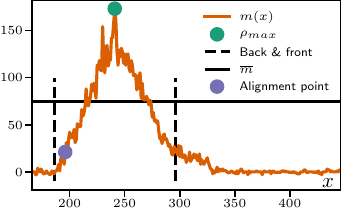}
  \caption{Sketch of the procedure to align flocks of the same age as
    described in the text.}
  \label{fig:align}
\end{figure}

We used the rear edge of the flocks to align them. To do so, consider
the case of a flock with positive magnetization shown in
Fig.~\ref{fig:align}. We first localize the site $i_{\rm max}$ where
the density is maximal. Going backward from there, $i_{\rm back}$ is
the first site where the magnetization changes sign. We then calculate $\bar m$ the
average magnetization per site over the region of size
$2\ell+1$ centered on $i_{\rm max}$, where $\ell\equiv |i_{\rm max}-i_{\rm back}|$.
Starting from $i_{\rm back}$, we then find the first site
$i_{\rm align}$ such that $m_{i_{\rm align}}>\bar m/4$. All
right-going flocks are then aligned such that their site
$i_{\rm align}$ coincide. Left-going flocks are transformed by
$m_i\to -m_i$ and $i \to L-i$ and then run through the same algorithm. 

\section{Another zero-temperature limit}
\label{Sec:0Talpha}

Considering the AIM at finite temperature and taking
the $\beta\to\infty$ limit lead to a zero-temperature model slightly
different from the one treated in Section~\ref{sec:asters}. Indeed, it
becomes possible to achieve configurations with vanishing
magnetization but finite density. These are such that $n^+_i=n_i^-\neq 0$, so that their
escape rates remain finite as $\beta\to \infty$ since
$W(s\to -s|m_i=0)=\omega_0$ for any $\beta$. Adding or removing one
particle, however, leads to a non-vanishing magnetization ---and hence an
immediate spin flip--- so that the only new configuration that may play
a role is $n^+_i=n_i^-=1$.

\begin{figure}[t]
  \centering
  \includegraphics[width=0.95\columnwidth]{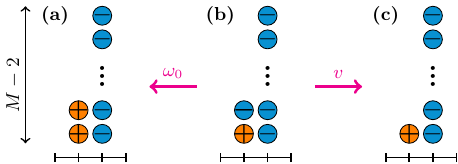}  
  \caption{The zero temperature limit of the AIM admits a new
    configuration with finite escape rate, shown in {\bf (b)}, where
    one site can be occupied by two particles with opposite signs,
    while the second site is occupied by $\na-2$
    particles. All moves lead to the evaporation of the aster but two:
    the minus particle on the first site can flip, at rate $\omega_0$,
    to reach the configuration $(2,\na-2)$ depicted in {\bf (a)}; the
    minus particle on the first site can hop, at rate $v$, to reach
    the configuration $(1,\na-2)$ depicted in {\bf (c)}.}\label{fig:newconf}
\end{figure}

\begin{figure}
  \includegraphics[width=.48\columnwidth]{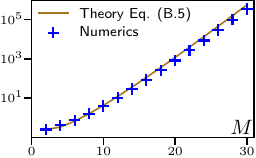} \includegraphics[width=.48\columnwidth]{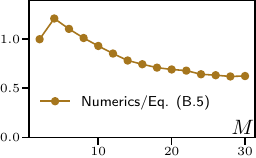}
  \caption{{\bf Left:} Numerical measurements of the MFPT till
    evaporation of an aster in the zero-temperature limit of the AIM,
    compared to the prediction~\eqref{eq:Tdaimscaling}. {\bf Right:}
    Taking the ratio of the data and the prediction
    ~\eqref{eq:Tdaimscaling} shows the convergence to a constant
    slightly different from $1$ as $M\to\infty$, due to the
    approximation involved in replacing $T_{2,\na-2}$ by its former
    expression~\eqref{eq:Tnapprox}.}\label{fig:data0Taim}
\end{figure}

Let us consider the mean first-passage time starting from an aster
with $\tfrac{\na}{2}$ spin $+$ particles at one site and $\tfrac{\na}{2}$ spin $-$ particles at
the next site. The dynamics is then identical to that described in
Section~\ref{sec:asters} until the aster reaches the configuration
$(1,\na-1)$ (or the symmetric case). Then, at rate $v(\na-1)$, it
reaches a new configuration, depicted in Fig.~\ref{fig:newconf}-b,
where the magnetization vanishes on the first site. This leads to the
recurrence
\begin{equation}\label{eq:Td1}
  T_{1,\na-1}=\frac{1}{r}+\mathfrak{q}\,T_{d,\na-2}\;,
\end{equation}
where $r=v \na$ is the total escape rate out of the configuration,
$\mathfrak{q}=\frac{\na-1}{\na}$ is the probability that one of the $\na-1$
spin $-$ particles hops forward during the next configuration change,
and $T_{d,\na-2}$ is the mean first passage time to evaporation starting from the
configuration shown in Fig.~\ref{fig:newconf}-b. Inspection of the dynamics starting from this
configuration then leads to
\begin{equation}\label{eq:Tdd}
  T_{d,\na-2}=\frac{1}{r_d}+\frac{v}{r_d}\,T_{1,\na-2}+\frac{\omega_0}{r_d}\,T_{2,\na-2}\;,
\end{equation}
where $r_d=2\omega_0+v\na$ is the escape rate. Injecting
Eq.~\eqref{eq:Tdd} into Eq.~\eqref{eq:Td1} then leads to:
\begin{equation}\label{eq:Tdfinal}
  T_{1,\na-1}=\frac{1}{r}+\frac{\mathfrak{q}}{r_d}+ \frac{\mathfrak{q}v}{r_d}\,T_{1,\na-2}+\frac{\mathfrak{q}\omega_0}{ r_d}\,T_{2,\na-2}\;.
\end{equation}
This has to be compared with the case considered in Section~\ref{sec:asters}, which reads
\begin{equation}\label{eq:Td1aim}
  T_{1,\na-1}=\frac{1}{r}+ \mathfrak{q}v T_{2,\na-2}\;.
\end{equation}

Two comments are in order. First, Eq.~\eqref{eq:Tdfinal} couples the
mean first passage time to evaporation with $\na$ and $\na-1$
particles, which makes an exact solution harder to find. Second, in
Eq.~\eqref{eq:Td1aim} $T_{1,\na-1}\sim T_{2,\na-2}$ as
$\na\to\infty$. On the contrary, in Eq.~\eqref{eq:Tdfinal}, there is
an extra factor $r_d^{-1}\sim 1/\na$ between $T_{1,\na-1}$ and
$T_{2,\na-2}$. The new configuration thus makes the evaporation
happens $\na$ times faster. Since the dynamics close to $\na/2$ is
insensitive to the details of the dynamics close to the evaporation,
it is natural to approximate $T_{2,\na-2}$ by its original
expression~\eqref{eq:Tnapprox}. This then leads to the self-consistent
scaling
\begin{equation}\label{eq:Tdaimscaling}
  T_{1,\na-1} \sim \frac{\omega_0}{2\omega_0+v \na }\frac{2^\na}{2 v \na}\;.
\end{equation}
As shown in Fig.~\ref{fig:data0Taim}, 
scaling~\eqref{eq:Tdaimscaling} agrees very well with our
numerics. The prefactor is, unsurprisingly, incorrect, due to the
approximation of $T_{2,\na-2}$ by its original
expression~\eqref{eq:Tnapprox}. All in all, the aster still has an
average life time that diverges exponentially with its number of
particles, and the physics remains qualitatively
unaltered from the case discussed in the main text.

\bibliography{refs.bib}

\end{document}